\documentclass[12pt]{article}
\usepackage[T1]{fontenc}
\usepackage[brazil,english]{babel}
\usepackage{amssymb,amsmath,amsbsy}
\usepackage[colorlinks=true]{hyperref}
\usepackage{graphicx}
    \graphicspath{{./Figures/}}
\usepackage{subfigure}
\usepackage{floatrow}
\usepackage{mathtools}
\usepackage[tmargin=3cm,lmargin=3cm,rmargin=2cm,bmargin=2cm]{geometry}
\usepackage{setspace}
\usepackage[square,numbers,sort&compress]{natbib}
\usepackage{longtable,makecell}
\usepackage{xcolor}
\usepackage[nopatch]{microtype}
\usepackage{multirow}
\usepackage{stix2} 
\usepackage{array}
\usepackage{caption}
\usepackage[normalem]{ulem} 
\usepackage{siunitx} 

\DeclareMathOperator{\sech}{sech} 
\newcommand{\sqabs}[1]{{\left|#1\right|}^2} 
\newcommand{\ZZ}{\mathbb{Z}} 
\newcommand{\RR}{\mathbb{R}} 
\newcommand{\CC}{\mathbb{C}} 
\newcommand{\II}{\mathbb{I}} 
\newcommand{\hypergeom}[1]{{}_2F_1\!\left(#1\right)} 
\newcommand{\diff}[2]{\frac{\text{d}#1}{\text{d}#2}} 
\newcommand{\ddiff}[2]{\frac{\text{d}^2#1}{{\text{d}#2}^2}} 

\title{\bf{Three-dimensional bound states of cylindrical quantum heterostructures with  position-dependent mass carriers}
\footnote{corresponding author: hugo.christiansen@ifce.edu.br

\,\,\,\ This version published in Physica Scripta, \textit{Phys. Scr.} 99 (2024) 015915

\,\,\,\ https://doi.org/10.1088/1402-4896/ad11c1}

\author{H. R. Christiansen$^{\dagger}$\, {\normalsize{and}} R. M. Lima$^{§} $ 
  \\
  \footnotesize{$^{\dagger}$IFCE -\, Instituto Federal de Educação, Ciência e Tecnologia do Ceará, CEP 61940-750, CE, Brazil} 
  \\
 \footnotesize{$^{§}$CBPF -\, Centro Brasileiro de Pesquisas Físicas, CEP 22290-180, Rio de Janeiro, RJ, Brazil} 
}
}

\date{}

\begin{document}
\maketitle

\begin{abstract}
We present a comprehensive spectral analysis of cylindrical quantum heterostructures by considering effective electronic carriers with position-dependent mass for five different kinetic-operator orderings. 
We obtain the bound energy eigenstates of particles in a three-dimensional cylindrical nanowire under a confining hyperbolic potential with both open and closed boundary conditions in the radial and the axial directions. In the present model we consider carriers with continuous mass distributions within the dot with abrupt mass discontinuities at the barriers, moving in a quantum dot that connects different substances. Continuity of mass and potential at the interfaces with the external layers result as a particular case. Our approach is mostly analytical and allows a precise comparison among von Roos ordering classes.\vspace{.35cm}
\end{abstract}


\section{Introduction}

The idea of considering quantum systems with position-dependent effective mass  (PDM) carriers initially arose with the development of solid state physics \cite{wannier:1937, slater:1949, luttinger:kohn:1955, bendaniel:duke:1966, gora:williams:1969}, more specifically of the envelope function approximation within semiconductor theory \cite{vonroos:1983, bastard:1992}. 
{The effective mass approximation can be understood as a renormalization of the electron mass in response to an external perturbation in a quantum heterostructure \cite{whalley:etal:2019}. For example, an electron coupled to collective excitations can be considered in order to perform band structure calculations of significant material properties in photovoltaic semiconductor devices.  Among them, optical properties as exciton binding energies, defect properties as donor acceptor energy levels, and transport properties (e.g. polaron radii and carrier mobility) can be computed from first principles.   Several effective mass definitions are possible among which the curvature or conductivity effective mass is the standard one. Its name is due for it is inversely proportional to the curvature of the electronic dispersion in momentum space and can be calculated from the band structure as \(m = \hbar^2 (\partial^2E/\partial k^2)^{-1}\). For a parabolic curvature band dispersion the charge carrier energy \(E \sim k^2\) and thus the curvature and the effective mass are constants. 
For this curvature the transport effective mass is equal to the curvature effective mass, but these are valid only for a region close to the conduction band minimum. For nonparabolic curvature band dispersions
one has to keep the second-order ellipsoidal energy surfaces and
introduce a nonlinear dependence on the energy. The transport effective mas results \(m = m_0 (1+2\alpha E)\),
where $m_0$ is the transport effective mass at the band edge and \(\alpha\) characterizes the nonparabolicity as the eigenenergy separates from the band edge. The bands can be increasingly filled so that the effective mass will also increase. This carrier concentration may depend on position along the device through selective photoexcitation and doping or temperature gradients, so the effective mass will vary correspondingly  
\cite{whalley:etal:2019, feneberg:etal:2016}.
Crystal structures of conventional materials (e.g. zeolites) as well as exotic ones (e.g. fullerenes nanotubes)  can be modeled as curved surfaces if one smears the atoms on a surface in the effective mass sense. It is a fundamental issue to calculate the mobility of electronic carriers over such surfaces. In this case one can use a Schrodinger equation with a curved metric to obtain the band structure with an effective mass. In case of adopting topological foams of graphite to realize curved surfaces, then the equation of motion of quasiparticles on the network of the honeycomb lattice will become, in the effective-mass picture, the problem of a zero mass Dirac equation (i.e., Weyl’s equation) on curved surfaces. 
\cite{aoki:etal:2001}}

 {In the effective mass approach, the renormalization of mass is treated in connection to the fermion-phonon coupling. The consideration of nonuniform effective masses is mathematically complicated and for this reason less frequent in this kind of investigation. However,  there are some interesting papers in this respect.
For example, in \cite{zhao:liang:ban:2003} a variational method is applied to obtain the dependence of the eigenenergies of bound polarons in parabolic quantum wells on the alloy composition   {\(\text{GaAs}/\text{Al}_\chi\text{Ga}_{1-\chi}\text{As}\)}. From numerical results it is shown that there is a clear contribution of the interaction between the electron and the
longitudinal optical phonons and the interphase phonons on the polaron energy levels. The position-dependent effective mass is considered to vary along the axis within the quantum well and is constant in the barrier material. We will do the same in the model we will present.
Particle-phonon coupling effects have been also addressed in the study of nuclei interactions. In \cite{saperstein:etal:2016} renormalization of the ends coming from pole phonon coupling contribution to the nucleon mass is computed. Calculations are done within a many-body Skyrme–Hartree Fock method, which makes the effective mass to be coordinate-dependent and strongly different from the bare one.  
Another interesting case is presented in \cite{fernandez-ramos:smedarchina:siebrand:2014} where a multidimensional Hamiltonian is developed for the investigation of tunneling dynamics in molecular systems. In order to prepare it for diagonalization, it is transformed into a lower dimension Hamiltonian by incorporating modes that move faster than the tunneling. This dimensional reduction leads to a mass renormalization manifesting through a coordinate-dependent mass kinetic energy operator. For this term, a Hermitian form of the von Roof class  is chosen and tested for the stability of the energy eigenvalues. }

Among the broad spectrum of applications \cite{cunha:christiansen:2013, christiansen:cunha:2013, christiansen:cunha:2014, dacosta:gomez:portesi:2020, ho:roy:2019, schmidt:dejesus:2018}, the study of heterostructures has been one particularly standing out \cite{elnabulsi:2020,valencia-torres:etal:2020, ganguly:etal:2006, galbraith:duggan:1988}. Quantum dots (microscopic structures in which the electrons dynamics are strongly restricted in space) can be modelled by considering charge carriers with position-dependent masses \cite{sari:etal:2019, elnabulsi:2020,elnabulsi:2020b, kasapoglu:etal:2021, kasapoglu:duque:2021}. The electronic and optical properties of these nanostructures can be engineered and fine tuned by changing their size and shape \cite{talapin:etal:2010, shirasaki:etal:2013, khordad:etal:2011, hassanabadi:rajabi:2009,lima:christiansen:2023b}. It is often useful to make nanowire connections of cylindrical shape \cite{kasapoglu:etal:2010, atayan:etal:2008}. In this geometry it is possible to control the energy spectrum with just two structure parameters such as the radius and length of the cylinder.

Quantum dots posses convenient optical features to modulate and control the output intensity of a component device. For instance, one can manipulate the electronic energy of such structures and adjust the absorption threshold frequency by regulating its growth.
 {Single-electron transistors can be used to study electron tunneling through a system of two tunnel junctions in series. An example of this is the observation of Coulomb-blockade oscillations of the conductance in a  quantum heterostructure as the voltage on a nearby gate electrode is varied. In \cite{kuo:chang:2003} this tunneling current is studied in a In-GaAs/GaAs quantum dot two energy-level model irradiated by infrared light within an effective position-dependent mass model. }

 {The absorption coefficients and relative changes in the refractive index for the inter sub-band transitions between the lower-lying allowed energy levels in cylindrical quantum dots have been investigated for different hyperbolic-type axial potentials. 
Diverse studies of cylindrical quantum dots have been performed considering hyperbolic Poschl-Teller type potentials \cite{wen-fang:2006, mora-ramos:barseghyan:duque:2010, hayrapetyan:kazaryan:tevosyan:2013, hayrapetyan:kazaryan:tevosyan:2012}. Hyperbolic potentials are well known for their applications in several branches of physics, including molecular and solid state physics. PDM quantum mechanics with hyperbolic potentials classes in one dimension has been treated analytically in \cite{christiansen:cunha:2013, christiansen:cunha:2014}.} In the present effort, we will add an external hyperbolic potential to a three-dimensional PDM kinetic Hamiltonian. The investigation of the shape effect of the quantum dots under external fields on the electronic spectrum and optical responses plays an important role in semiconductor physics. It helps to simulate real situations and improve the performance of optoelectronic equipments based on low-dimensional heterostructures.

 {The general Hermitian Hamiltonian for a quantum carrier with non-uniform mass 
\(M \equiv M(\boldsymbol{r})\) can be written as 
\[\hat{H}_{a,b}(\boldsymbol{r}) = \frac{1}{4} \left( M^a \hat{\boldsymbol{p}} M^{-1-a-b} \hat{\boldsymbol{p}} M^b + M^b \hat{\boldsymbol{p}} M^{-1-a-b} \hat{\boldsymbol{p}} M^a \right) + V(\boldsymbol{r}) \; ,\]
(see von Roos \cite{vonroos:1983}), where 
\(\hat{\boldsymbol{p}} \equiv -i \hbar \boldsymbol{\nabla}\) is the linear momentum operator}, 
  \(a,b \in \RR\)\footnote{ {We will adopt \(\ZZ\), \(\RR\), \(\CC\) and \(\II\) to denote the sets of the integers, real, complex and purely imaginary numbers, respectively. $*$ superscript indicates that zero is excluded. $+$ and $-$ subscript indicates respectively only the positive and negative numbers.}} are the mass
 { or ordering parameters and \(V(\boldsymbol{r})\) is an external potential}.
 All frequently appearing Hamiltonians in the literature are particular cases of this one and are the usual alternative kinetic orderings for building energy operators. These special Hamiltonians will be denoted here by BDD (BenDaniel \& Duke) \cite{bendaniel:duke:1966} ($a=b=0$), GW (Gora \& Williams) \cite{gora:williams:1969} ($a=-1$, $b=0$), ZK (Zhu \& Kroemer) \cite{zhu:kroemer:1983} ($a=b=-1/2$), LK (Li \& Kuhn) \cite{li:kuhn:1993} ($a=0$, $b=-1/2$) and MM (Mustafa \& Mazharimousavi) \cite{mustafa:mazharimousavi:2007} ($a=b=-1/4$). 

These Hamiltonians can be studied simultaneously in order to find its solutions. For a continuous PDM, there are no general restrictions regarding the choice of the ordering and the boundary conditions are given as usual by the continuity of the wave-function and its derivative. On the other hand, for abrupt heterojunctions with finite discontinuities in the mass profile \cite{morrow:brownstein:1984, pistol:1999, thomsen:einevoll:hemmer:1989, smagley:mojahedi:osinski:2002} some considerations are in order. We will discuss them in the next sections.

In this work we will analyse a system representing a generic quantum wire, a nanometric double heterostructure with cylindrical symmetry. We will solve it with a general approach where the best established kinetic orderings are considered, making thus possible an assessment of the so called ambiguity problem of the PDM Hamiltonian. The bound-state energy spectrum of a particle with PDM subject to an external potential varying significantly in the axial direction will be analytically obtained. In the calculations of the eigenvalue equation we consider both the Dirichlet zero flux boundary conditions and the open non-zero flux boundary conditions which guarantees the validity of the results for quasi-steady states with extremely high lifetimes. The paper is organized as follows. In the next \hyperref[sec:gse]{Section}, we start by writing the Schrödinger equation with the von Roos Hamiltonian as a general dimensionless expression. In Section~\ref{sec:mcdh} we present the model of a double heterostructure which we will treat. In Subsection~\ref{subsec:sv:aep} we separate variables by taking into account the PDM, and thus we get the resulting potential. Then, we proceed to solve the angular and radial sectors and set up the boundary conditions for the axial solution in \ref{subsec:ras:bcas}. Section~\ref{sec:as} is dedicated to the axial differential equation and is divided into five subsections. In \ref{subsec:cpt} we perform a canonical point transformation in order to recast the initial differential equation as an ordinary Schrödinger equation. In \ref{subsec:sir} we obtain the analytical solutions whose energy spectrum is analysed in \ref{subsec:es} and plotted in \ref{subsec:gas}. Finally in Subsection~\ref{subsec:scpdcm} we discuss the continuous PDM special case. In Section~\ref{sec:e} we discuss the complete three dimensional eigenstates and Section~\ref{sec:c} is where we draw our conclusions.


\section{Generalized Schrödinger equation in 3D}
\label{sec:gse}

 {In three-dimensional space,  the position-dependent mass kinetic sector} can be rewritten as
\[\hat{T}_{a,b}(\boldsymbol{r})=\frac{1}{2M}\hat{\boldsymbol{p}}^2-\frac{1}{2M}\left(\frac{1}{M}\hat{\boldsymbol{p}}M\right)\cdot\hat{\boldsymbol{p}}+U_{a,b}(\boldsymbol{r})\;,\]
where
\[U_{a,b}(\boldsymbol{r}) \equiv \frac{1}{2M}\left[\frac{a+b}{2}\hat{\boldsymbol{p}}\left(\frac{1}{M}\hat{\boldsymbol{p}}M\right)-\left(ab+\frac{a+b}{2}\right){\left(\frac{1}{M}\hat{\boldsymbol{p}}M\right)}^2\right]\]
behaves like a potential of kinematic origins which depends on the initial operator ordering.

For a particle with total energy $E$ in an external potential \(V(\boldsymbol{r})\) the wave function \(\Psi(\boldsymbol{r})\) obeys a generalized Schrödinger equation (GSE)
\begin{multline*}\label{eq:0}
	-\frac{\hbar^2}{2M}{\boldsymbol{\nabla}}^2\Psi(\boldsymbol{r})+\frac{\hbar^2}{2M}\frac{\boldsymbol{\nabla}M}{M}\cdot\boldsymbol{\nabla}\Psi(\boldsymbol{r})+\left(U_{a,b}(\boldsymbol{r})+V(\boldsymbol{r})\right)\Psi(\boldsymbol{r})=E\Psi(\boldsymbol{r})\;,\\
	U_{a,b}(\boldsymbol{r})=-\frac{\hbar^2}{2M}\left[\frac{a+b}{2}\boldsymbol{\nabla}\cdot\left(\frac{\boldsymbol{\nabla}M}{M}\right)-\left(ab+\frac{a+b}{2}\right){\left(\frac{\boldsymbol{\nabla}M}{M}\right)}^2\right]\;.
\end{multline*}

By means of the following transformations: \(\boldsymbol{r} \to \epsilon\boldsymbol{r}\), \(\Psi(\boldsymbol{r}) \to \epsilon^{-D/2}\psi(\boldsymbol{r})\) and \(M(\boldsymbol{r}) \to m_0m(\boldsymbol{r})\), we can obtain a dimensionless rescaled version of the equation above. The coordinate \(\boldsymbol{r}\) is now dimensionless, $D$ is the number of spatial dimensions, \(\epsilon\) and $m_0$ are dimensional positive constants, and \(\psi(\boldsymbol{r})\) and \(m(\boldsymbol{r})\) are respectively the wave function and the mass, both dimensionless and rescaled. With this, the GSE above becomes
\begin{equation}
    \label{eq:1}
	-\frac{1}{m}{\boldsymbol{\nabla}}^2\psi\!\left(\boldsymbol{r}\right)+\frac{1}{m}\frac{\boldsymbol{\nabla}m}{m}\cdot\boldsymbol{\nabla}\psi\!\left(\boldsymbol{r}\right)+\left(\tilde{U}_{a,b}\!\left(\boldsymbol{r}\right)+\tilde{V}\!\left(\boldsymbol{r}\right)\right)\psi\!\left(\boldsymbol{r}\right)=\tilde{E}\psi\!\left(\boldsymbol{r}\right)\;,
\end{equation}
where
\begin{multline*}
    m \equiv m(\boldsymbol{r}) \;,\quad \tilde{E} \equiv \frac{2\epsilon^2m_0}{\hbar^2}E \;,\quad \tilde{V}\!\left(\boldsymbol{r}\right) \equiv \frac{2\epsilon^2m_0}{\hbar^2}V\!\left(\epsilon\boldsymbol{r}\right)
     \quad \text{and}\\
     \\
	\tilde{U}_{a,b}\!\left(\boldsymbol{r}\right) \equiv \frac{2\epsilon^2m_0}{\hbar^2}U_{a,b}\!\left(\epsilon\boldsymbol{r}\right)=-\frac{1}{m}\left[\frac{a+b}{2}\boldsymbol{\nabla}\cdot\left(\frac{\boldsymbol{\nabla}m}{m}\right)-\left(ab+\frac{a+b}{2}\right){\left(\frac{\boldsymbol{\nabla}m}{m}\right)}^2\right] \;.
\end{multline*}
 {Note that \(\tilde{E}\), \(\tilde{V}\!\left(\boldsymbol{r}\right)\) and \(\tilde{U}_{a,b}\!\left(\boldsymbol{r}\right)\) are dimensionless, as \(\psi(\boldsymbol{r})\) and \(m(\boldsymbol{r})\). The (dimensional) energy spectrum, $E$, is proportional to \(\tilde{E}\) through the dimensional factor \(\frac{\hbar^2}{2\epsilon^2m_0}\). The interest in doing so is that in this way the parameters $m_0$ and \(\epsilon\), controlling  the electronic mass and the size of the quantum dot, can be ignored until the end of the whole calculation. We will further comment on this later.}

For continuous mass functions, the wave-function and its derivative have to be continuous along the whole heterostructure. When there are abrupt heterojunctions and finite discontinuities in the mass profile \cite{morrow:brownstein:1984, pistol:1999, thomsen:einevoll:hemmer:1989, smagley:mojahedi:osinski:2002} we have to observe the ordering class. For the orderings where $b=a$, the boundary conditions are set by the continuity of \(M^a\Psi(\boldsymbol{r})\) and \(M^{-(1+a)}\boldsymbol{\nabla}_{\hat{\boldsymbol{n}}}\Psi(\boldsymbol{r})\), where \(\boldsymbol{\nabla}_{\hat{\boldsymbol{n}}}\) denotes the directional derivative with respect to the unit vector \(\boldsymbol{n}\) normal to the interface of discontinuity. On the other hand, the orderings where \(b\neq a\) (this is the case of the GW and LK Hamiltonians) must be discarded for two reasons: (i) As shown in \cite{morrow:brownstein:1984} the wave function vanishes at the interfaces; it implies that there is no probability current across the interfaces so the discontinuities act as infinite barriers. (ii) The energy of the fundamental state diverges negatively \cite{thomsen:einevoll:hemmer:1989}. These two aspects characterize a non-physical situation in the present problem and must be avoided. 

\section{Model of cylindrical double-heterostructure}
\label{sec:mcdh}

The theoretical treatment of cylindrical quantum dots is usually considered with an infinite confinement potential, either in the axial direction or in the radial direction. This facilitates the analytical decoupling of the eigenvalue differential equation. The infinite potential approach is applicable when the physical system corresponds, for example, to a quantum wire made of a semiconductor surrounded by a vacuum or by a different material with significantly higher energy gap. Separation of variables with finite confinement potentials is also viable provided the electron wave-function is mostly confined to the quantum dot domain. However, if the dimension of the nanocrystal is similar to the effective Bohr radius of the material the approximation is not applicable. Assuming a quantum semiconductor surrounded by a material inducing a finite potential barrier, it is usual to resort to numerical methods where the differential operators are discretized in space. This is particularly common when effects such as spatial variation of the effective mass associated with different semiconductor materials are included.

Here, in a quite comprehensive approach, we will consider a model of a solid such as a quantum nanowire, where a particle is constrained to a cylindrical region of radius \(\epsilon\delta_r\), \(\delta_r>0\), around the $x$ axis of a Cartesian coordinate system. Inside, the mass of the particle and the external potential are functions of the $x$ coordinate, \(M=M(x) \Leftrightarrow m=m(x)\) and \(V(\boldsymbol{r})=V(x) \Leftrightarrow \tilde{V}(\boldsymbol{r})=\tilde{V}(x)\).

We will pursuit the case of a double heterostructure where the mass and the external potential are given by
\begin{equation}
    \label{eq:2}
    M(x)=
    \begin{cases}
        m_0m_1              &   x \leq -\epsilon\delta_x \\
        m_0m_\text{in}{\left(\frac{x}{\epsilon}\right)}   &   |x|<\epsilon\delta_x \\
        m_0m_2              &   x \geq \epsilon\delta_x
    \end{cases} \quad \text{and} \quad 
    V(x)=
    \begin{cases} 0       &   |x| \geq \epsilon\delta_x \\
        \dfrac{\hbar^2}{2\epsilon^2m_0}\tilde{V}_\text{in}{\left(\frac{x}{\epsilon}\right)}   &   |x|<\epsilon\delta_x
    \end{cases} \;.
\end{equation}
 {The subindex \({}_\text{in}\) signals the value in the Intermediate Region {(IR)}.}
 {Within the quantum dot, in the interval \(|x|<\epsilon\delta_x\), \(\delta_x>0\), we assume a mass and a confining potential varying along the cylinder axis and radially uniform. This is a common assumption for quantum dots with azimuthal symmetry and is consistent with the standard dot growth technique. An infinite barrier at the external radius of the cylindrical configuration is also fitting for a quantum nanostructure surrounded by a vacuum in the radial direction. Additionally, it guarantees a clean separation between the axial and radial variables in the nontrivial differential  {eq.~\eqref{eq:1}} which one aims to solve. For example, in \cite{zhao:liang:ban:2003} the authors obtain the dependence of the eigenenergies of bound polarons in quantum wells on the alloy composition  {\(\text{GaAs}/\text{Al}_\chi\text{Ga}_{1-\chi}\text{As}\)}. There, the position-dependent effective mass is considered to vary along the axis within the quantum well and is uniform in the barrier materials. Outside the quantum well,  the potential is also set constant. In \cite{kasapoglu:etal:2022} this assumption allows a theoretical analysis of the inter-subband absorption coefficients for a hyperbolic-type PDM carrier confined within potentials of a family of hyperbolic functions previously studied in \cite{christiansen:cunha:2013}  for solitonic position dependent mass in one dimensional quantum mechanics. Considering  {\(\text{GaAs}/\text{Al}_\chi\text{Ga}_{1-\chi}\text{As}\)} a direct connection is established between the electron effective mass and the gap energy of the semiconductor which depends on the aluminum concentration. In turn, it can be related with the coordinate dependence of the confinement potential. } 

 {It is then appropriate to adopt a solitonic mass profile along the cylinder axis,
\begin{equation}
    \label{eq:3}
    m_\text{in}(x)=\sech^2{x},
\end{equation}
and for the external potential taking a confining hyperbolic  well \(\tilde{V}_\text{in}(x)=\tilde{V}_0\left(1-\frac{\sinh^2{x}}{\sinh^2{\delta_x}}\right)\), both shapes relevant in several branches of physics. This kind of potential wells can be easily generated and controlled experimentally. A growth process of the alloy  {\(\text{GaAs}/\text{Al}_\chi\text{Ga}_{1-\chi}\text{As}\)} can be set up by adjusting the concentration of aluminum to obtain the potential profile. The local droplet epitaxy technique allows the development of such heterostructures with the desired symmetry and space distribution, see e.g. \cite{stemmann:etal:2009}. This effective mass density, which is functionally analogous to the potential,  is also convenient to control singularities in the differential equation.
For example, for the investigation of optical properties as the relative changes of the refraction index
coefficient in the intersubband transition between the lower-lying electronic levels of the Manning
potential in isolated quantum wells \cite{kasapoglu:duque:2021}. See also \cite{kasapoglu:yucel:duque:2023} where  quantum wells are studied adopting the solitonic mass  combined with a Rosen-Morse potential.
Interestingly, in a recent paper \cite{zhang:etal:2023}  soliton states of electrons with strong quadratic coupling to the phonon coordinates have been found in an exact numerical study.  This polaron model, namely a system of electrons renormalized by their coupling to lattice vibrations soft phonons, shows a non-monotonic dependence of the soliton effective mass and the residue.  This choice of mass profile and potential is also in accordance to previous mathematical efforts \cite{cunha:christiansen:2013, christiansen:cunha:2013, christiansen:cunha:2014,bagchi:etal:2004}\footnote{For a comprehensive study on alternative position-dependent mass in one dimensional Hamiltonians, see \cite{lima:christiansen:2022}.}.  } 

 {The values of the system parameters in the external layers $m_1$, $m_2$, \(\delta_x\), \(\delta_r\) and \(\tilde{V}_0\) can be conveniently adjusted depending on the system characteristics (e.g., \(\tilde{V}_0\)  can be experimentally controlled, whereas $m_1$ and $m_2$ are determined by the composition of the material). For simplicity, we assume $m_1>m_2$ and \(\tilde{V}_0<0\). In view of the scale transformation \(\boldsymbol{r} \to \epsilon\boldsymbol{r}\) made in Sec.~\ref{sec:gse}, we are free to set \(\delta_r=1\), as long as the scale parameter \(\epsilon\) takes the appropriate values (see Sec.~\ref{sec:c}).}


\subsection{The axial effective potential in the cylinder}
\label{subsec:sv:aep}

We solve eq.~\eqref{eq:1} in cylindrical coordinates,
\((x,y,z)=(x,r\cos{\phi},r\sin{\phi}), x \in \RR, r>0, 0 \leq \phi \leq 2\pi,\)
and assume \(\psi(\boldsymbol{r})=R(r)\Phi(\phi)X(x)\), where $R(r)$, \(\Phi(\phi)\) and $X(x)$ are respectively the radial, azimuthal and axial functions. We thus obtain three equations
\begin{subequations}
    \label{eq:4}
    \begin{equation}
        \label{eq:4a}
        \Phi''(\phi)+\mathscr{k}^2\Phi(\phi)=0 \;,\quad \mathscr{k} \in \CC \;,
    \end{equation}
    \begin{equation}
        \label{eq:4b}
        r^2R''(r)+rR'(r)+\left(\mathscr{p}^2r^2-\mathscr{k}^2\right)R(r)=0 \;,\quad \mathscr{p} \in \CC \;,
    \end{equation}
    \begin{equation}
        \label{eq:4c}
        -\frac{1}{m}X''(x)+\frac{m'}{m^2}X'(x)+\left(\tilde{U}_{a,b}(x)+\tilde{V}_\mathscr{p}(x)\right)X(x)=\tilde{E}X(x)\;,
    \end{equation}
\end{subequations}
where
\[\tilde{U}_{a,b}(x)=-\frac{1}{m}\left[\frac{a+b}{2}{\left(\frac{m'}{m}\right)}'-\left(ab+\frac{a+b}{2}\right){\left(\frac{m'}{m}\right)}^2\right]\]
is the general ordering kinetic potential shown in Sec.\ref{sec:gse} (prime in this expression denotes derivative with respect to $x$) which is added to another mass dependent potential term
\[\tilde{V}_\mathscr{p}(x) \equiv \tilde{V}(x)+\frac{\mathscr{p}^2}{m}=
\begin{cases}
    \dfrac{\mathscr{p}^2}{m_1}                                                                              &   x \leq -\delta_x \\
    \tilde{V}_0+\mathscr{p}^2+\left(\mathscr{p}^2-\dfrac{\tilde{V}_0}{\sinh^2{\delta_x}}\right)\sinh^2{x}   &   |x|<\delta_x \\
    \dfrac{\mathscr{p}^2}{m_2}                                                                              &   x \geq \delta_x
\end{cases}\]
This is the \(\mathscr{p}\)-depending \emph{axial effective potential} (the inverse-mass term arose from the dynamics in the perpendicular space and adds up along the whole line). Later we will see that \(\mathscr{p}\) is a real positive number, \(\mathscr{p}>0\), which ensures that \(\tilde{V}_\mathscr{p}(x) \in \RR\). 

\begin{figure}[H]
	\subfigure[\label{fig:1a}]{\includegraphics[width=.49\linewidth]{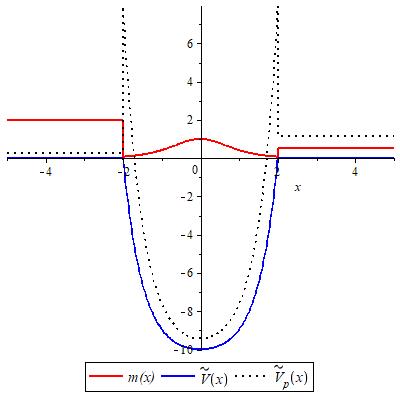}} \ \
	\subfigure[\label{fig:1b}]{\includegraphics[width=.49\linewidth]{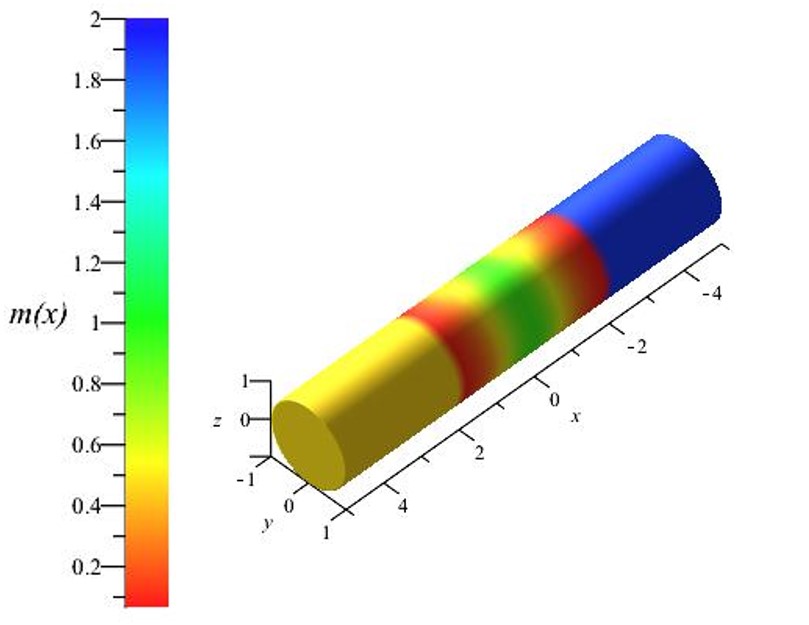}} 
\caption{Double-heterostructure with cylindrical symmetry. In \subref{fig:1a}, the red solid line is for the position dependent mass, the blue solid line is for the rescaled external potential and the dotted line is for the rescaled axial effective potential. The three-dimensional cylinder in figure \subref{fig:1b} represents the heterostructure where the colors depict the different values of the mass, as indicated. Here, $m_1=2$, $m_2=0.5$, \(\delta_x=2\), \(\tilde{V}_0=-10\) and \(\mathscr{p}=0.75\).  {The cylindrical quantum dot is the central region going from red to red, while the external layers are respectively yellow and blue.}}
	\label{fig:1}
\end{figure}

\begin{figure}[H]
	\subfigure[\label{fig:2a}]{\includegraphics[width=.49\linewidth]{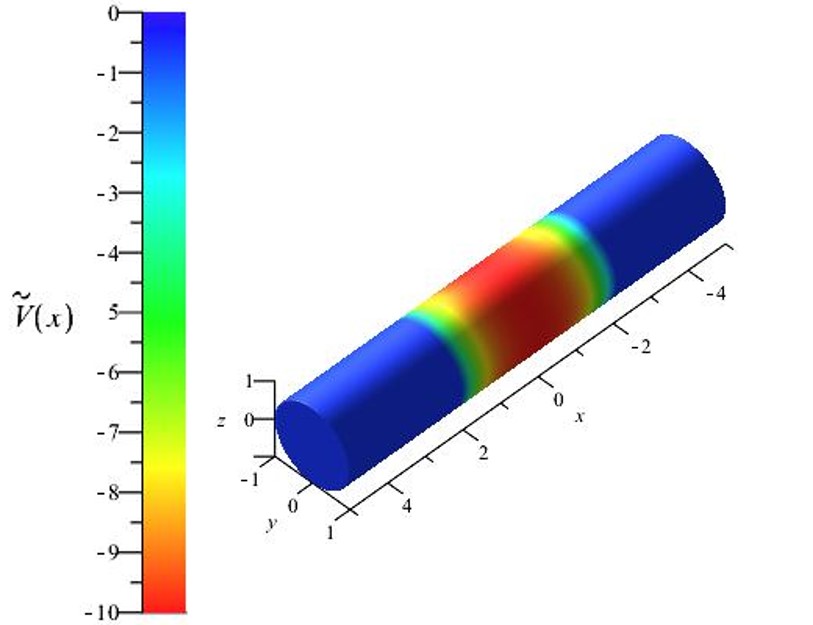}} \ \
	\subfigure[\label{fig:2b}]{\includegraphics[width=.49\linewidth]{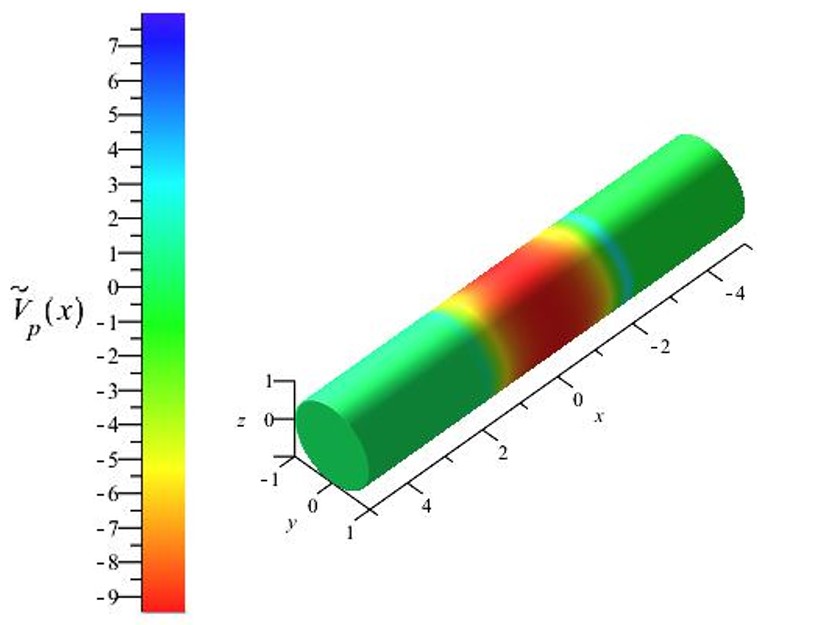}}
	\caption{Double-heterostructure with cylindrical symmetry.  The three-dimensional cylinders in figures  \subref{fig:2a} and \subref{fig:2b} represent the heterostructure where the colors depict the different values of rescaled potentials, as indicated. Here, $m_1=2$, $m_2=0.5$, \(\delta_x=2\), \(\tilde{V}_0=-10\) and \(\mathscr{p}=0.75\).  {The cylindrical quantum dot is the central region. In \subref{fig:2a} going from blue to blue external layers and in \subref{fig:2b} going from the left green to the right (different) green external layer.}}
	\label{fig:2}
\end{figure}

Since \(\mathscr{p}^2-\frac{\tilde{V}_0}{\sinh^2{\delta_x}}>0\), then a necessary and sufficient condition for the existence of bound states is \(\tilde{V}_0+\mathscr{p}^2<\frac{\mathscr{p}^2}{m_1}\), which is always obeyed when \(m_1 \leq 1\) (otherwise, one has the constraint \(\mathscr{p}^2<\frac{m_1}{1-m_1}\tilde{V}_0\)). In Figs.~\ref{fig:1}-~\ref{fig:2} we represent this system for some choice of the parameters.

\subsection{Radial and angular solutions and boundary conditions for the axial solution}
\label{subsec:ras:bcas}

Periodicity conditions on \(\Phi(\phi)\) are known to imply
\begin{equation}
    \label{eq:5}
    \Phi_\mathscr{k}(\phi)=A_\mathscr{k}e^{i\mathscr{k}\phi} \;,\quad \mathscr{k} \in \ZZ \;,\quad A_\mathscr{k} \in \CC \;,
\end{equation}
 {where the parameter \(\mathscr{k}\) corresponds to the quantum magnetic number, see eq.~\eqref{eq:4a}.}

Note that for each value of \(\mathscr{k}\) the solutions of eq.~\eqref{eq:4b}  must be finite at the origin and zero at \(r=\delta_r=1\).  {The quantum number
\(\mathscr{p}\) depends implicitly on \(\mathscr{k}\) and \(\mathscr{\ell}\) such that 
\(\mathscr{p} \equiv \mathscr{p}_{\mathscr{k},\ell}\), where \(\mathscr{p}_{\mathscr{k},\ell}\) (\(\ell \in \ZZ_+^*\)) is the \(\ell\)-th positive root of \(J_\mathscr{k}(\mathscr{p})=0\) }
\cite[p.~370]{abramowitz:stegun:1970}. 
The radial solutions thus read
\begin{equation}
    \label{eq:6}
    R_{\mathscr{k},\ell}(r)=B_{\mathscr{k},\ell}J_\mathscr{k}(\mathscr{p}_{\mathscr{k},\ell}r) \;,\quad \mathscr{k} \in \ZZ \;,\quad \ell \in \ZZ_+^* \;,\quad B_{\mathscr{k},\ell} \in \CC \;.
\end{equation}
The zeros of first kind Bessel functions are compactly related by \(\mathscr{p}_{\mathscr{k},\ell}<\mathscr{p}_{\mathscr{k}+1,\ell}<\mathscr{p}_{\mathscr{k},\ell+1}\) (see \cite[eq.~9.5.2]{abramowitz:stegun:1970} and \cite[p.~409]{abramowitz:stegun:1970}). Thus, including this in the discussion of Subsection~\ref{subsec:sv:aep}, for  $m_1>1$ there must be a maximum value \( \mathscr{k}_\text{max} \geq 0\) (where \(\mathscr{k}=-\mathscr{k}_\text{max},...,\mathscr{k}_\text{max}\)) and for each \(\mathscr{k}\) the values of \(\ell\) are also restricted to a maximum \(\ell_\text{max}>0\) (\(\ell=1,...,\ell_\text{max}\)).

Eqs.~\eqref{eq:4a} and \eqref{eq:4b} are thus the same as those obtained for constant mass. Eq.~\eqref{eq:4c}, on the other hand, tells us that for a PDM, the function $X(x)$ obeys a 1D-GSE with an effective potential (in the next \hyperref[sec:as]{Section} we analyze this equation in details).

Let us write the corresponding bound-state wavefunctions as \(X(x)=C\mathcal{X}(x)\), where \(C \in \CC\) results from normalization and \(\mathcal{X}(x)\) is a piecewise function
\begin{equation}
    \label{eq:7}
    \mathcal{X}(x)=
    \begin{cases}
        e^{\theta_1x}                                           &   x \leq -\delta_x \\
        c^{(1)}X_\text{in}^{(1)}(x)+c^{(2)}X_\text{in}^{(2)}(x) &   |x|<\delta_x \\
        ce^{-\theta_2x}                                         &   x \geq \delta_x
    \end{cases} \;,
\end{equation}
with \(\theta_\tau \equiv \sqrt{\left|m_\tau\tilde{E}-\mathscr{p}^2\right|}\) and \(c^{(\sigma)},c_\sigma \in \CC\) (\(\sigma,\tau=1,2\)). 

Boundary conditions for $X(x)$ are determined by continuity of $m^aX(x)$ and $m^{-(1+a)}X'(x)$, which lead us to the following expression for the allowed energy levels:
\begin{subequations}
    \label{eq:8}
    \begin{equation}
        \label{eq:8a}
        \begin{vmatrix}
            \kappa_1^{(1)}  &   \kappa_2^{(1)} \\
            \kappa_1^{(2)}  &   \kappa_2^{(2)}
        \end{vmatrix}
        =0 \;,\quad \text{where} \quad \kappa_\tau^{(\sigma)} \equiv {\left.\diff{X_\text{in}^{(\sigma)}}{x}\right|}_{x={(-1)}^\tau\delta_x}+{(-1)}^\tau{\left(\frac{\sech^2{\delta_x}}{m_\tau}\right)}^{2a+1}\theta_\tau X_\text{in}^{(\sigma)}\!\left({(-1)}^\tau\delta_x\right) \;.
    \end{equation}
    Considering the Wronskian of \(X_\text{in}^{(1)}(x)\) and \(X_\text{in}^{(2)}(x)\) evaluated at \(x={(-1)}^\tau\delta_x\),
    \[\mathcal{W}_\tau \equiv \mathcal{W}\!\left\{X_\text{in}^{(1)}\!\left({(-1)}^\tau\delta_x\right),X_\text{in}^{(2)}\!\left({(-1)}^\tau\delta_x\right)\right\}=
    {\begin{vmatrix}
        X_\text{in}^{(1)}(x)        &   X_\text{in}^{(2)}(x) \\
        \diff{X_\text{in}^{(1)}}{x} &   \diff{X_\text{in}^{(2)}}{x}
    \end{vmatrix}}_{x={(-1)}^\tau\delta_x} \;,\]
    and assuming that \(\mathcal{W}_1\neq0\), we can write the coefficients of eq.~\eqref{eq:7} as
    \begin{multline}
        \label{eq:8b}
        c^{(1)}={\left(\frac{\sech^2{\delta_x}}{m_1}\right)}^{-a}\frac{e^{-\theta_1\delta_x}\kappa_1^{(2)}}{\mathcal{W}_1} \;,\quad c^{(2)}=-{\left(\frac{\sech^2{\delta_x}}{m_1}\right)}^{-a}\frac{e^{-\theta_1\delta_x}\kappa_1^{(1)}}{\mathcal{W}_1} \;, \\ c={\left(\frac{m_2}{m_1}\right)}^{-a}\frac{e^{(\theta_2-\theta_1)\delta_x}}{\mathcal{W}_1}
        \begin{vmatrix}
            X_\text{in}^{(1)}(\delta_x) &   \kappa_1^{(1)} \\
            X_\text{in}^{(2)}(\delta_x) &   \kappa_1^{(2)}
        \end{vmatrix} \;.
    \end{multline}
\end{subequations}


\section{The axial solutions}
\label{sec:as}

Now, we proceed to calculate $X(x)$ to complete the total three-dimensional wave-function.


\subsection{Canonical point transformations}
\label{subsec:cpt}

In order to solve eq.~\eqref{eq:4c} in the IR we define a variable \(\xi\) and a function \(\Xi(\xi)\) such that
\begin{subequations}
    \label{eq:9}
    \begin{equation}
        \label{eq:9a}
        \diff{\xi}{x}=m_\text{in}^{1/2} \;,
    \end{equation}
    \begin{equation}
        \label{eq:9b}
        X_\text{in}(x)=m_\text{in}^{1/4}\Xi(\xi) \;.
    \end{equation}
\end{subequations}
With this, eq.~\eqref{eq:4c} transforms into
\begin{subequations}
    \label{eq:10}
    \begin{equation}
        \label{eq:10a}
        \Xi''(\xi)+\left(\tilde{E}-V_{a,b}(\xi)\right)\Xi(\xi)=0 \;,
    \end{equation}
    where
    \begin{equation}
        \label{eq:10b}
        V_{a,b}(\xi) \equiv \tilde{V}_\text{in}(x(\xi))+\frac{1}{m_\text{in}}\left[\mathscr{p}^2-\frac{1}{2}\left(a+b+\frac{1}{2}\right){\left(\frac{m_\text{in}'}{m_\text{in}}\right)}'+\left(ab+\frac{a+b}{2}+\frac{3}{16}\right){\left(\frac{m_\text{in}'}{m_\text{in}}\right)}^2\right]
    \end{equation}
\end{subequations}
is an effective potential in \(\xi\) space\footnote{The derivatives in eq.~\eqref{eq:10b} are taken with respect to $x$ and then expressed in terms of \(\xi\).} (\(\xi\)-EP).

Eq.~\eqref{eq:10a} is a typical Schrödinger equation for a constant-mass particle with energy eigenvalue \(\tilde{E}\) under the effective-potential \(V_{a,b}(\xi)\). This potential consists of the external potential rewritten in the new variable, \(\tilde{V}_\text{in}(x(\xi))\), then further deformed by the addition of a term which is different for each choice of the PDM plus kinetic-ordering set to solve the problem.

For \(|x|<\delta_x\), by combining eqs.~\eqref{eq:2} and \eqref{eq:9a} we get the variables transformation
\begin{equation}
    \label{eq:11}
    \cos{\xi}=\sech{x} \;,
\end{equation}
 mapping \((-\delta_x,\delta_x) \ni x \mapsto \xi \in \left(-\cos^{-1}{\sech{\delta_x}},\cos^{-1}{\sech{\delta_x}}\right)\). This, combined with eq.~\eqref{eq:10b}, gives the \(\xi\)-EP:
\[V_{a,b}(\xi)=\omega_{a,b}\tan^2{\xi}+V_{a,b}^{(0)} \;,\]
where
\[\omega_{a,b} \equiv \mathscr{p}^2-\frac{\tilde{V}_0}{\sinh^2{\delta_x}}+4ab+2(a+b)+\frac{3}{4} \quad \text{and} \quad V_{a,b}^{(0)} \equiv \tilde{V}_0+\mathscr{p}^2+a+b+\frac{1}{2} \;.\]


\subsection{Solutions in the Intermediate Region}
\label{subsec:sir}

Ignoring for a moment the mass parameters as indices of the eigenfunctions, we define a new variable \(w=\sin^2{\xi}\) and a function $W(w)$ such that
\[\Xi(\xi)=w^{\mu_\pm}{(1-w)}^{\nu_{a,b}}W(w) \;,\]
where \(\mu_\pm=\frac{1\pm1}{4}\) and \(\nu_{a,b}=\frac{1}{4}\left(1-\sqrt{1+4\omega_{a,b}}\right)\)~~
, then eq.~\eqref{eq:10a} results in a Gauss hypergeometric equation \cite[Chap.~15]{abramowitz:stegun:1970},
\[\left\{w(1-w)\ddiff{}{w}+\left[\gamma_\pm-(\alpha_{a,b}^\pm+\beta_{a,b}^\pm+1)w\right]\diff{}{w}-\alpha_{a,b}^\pm\beta_{a,b}^\pm\right\}W_{a,b}^\pm(w)=0 \;,\]
with coefficients 
{
\[\gamma_\pm=2\mu_\pm+\frac{1}{2}=1\pm\frac{1}{2}
\; , \;
\alpha_{a,b}^\pm=\mu_\pm+\nu_{a,b}+\frac{1}{2}\sqrt{\mathcal{E}_{a,b}+\omega_{a,b}}
\; \text{and} \;
\beta_{a,b}^\pm=\mu_\pm+\nu_{a,b}-\frac{1}{2}\sqrt{\mathcal{E}_{a,b}+\omega_{a,b}} \;.\]
where \(\mathcal{E}_{a,b} \equiv \tilde{E}-V_{a,b}^{(0)}\) is the energy measured with respect to the value of the \(\xi\)-EP at the origin.}

Since \(\gamma_\pm\notin\ZZ\), for any \(\mu\) the two linear independent solutions around $w=0$ are given in terms of Gauss hypergeometric functions: 
\(W_{a,b}^\pm(w)=\hypergeom{\alpha_{a,b}^\pm,\beta_{a,b}^\pm;\gamma_\pm;w}\) and 
\(\overline{W}_{a,b}^\pm(w)=w^{1-\gamma_\pm}\hypergeom{\alpha_{a,b}^\pm+1-\gamma_\pm,\beta_{a,b}^\pm+1-\gamma_\pm;2-\gamma_\pm;w}\). 
In \(\xi\) space these solutions read

\[\Xi_{a,b}^\pm(\xi)={\left(\sin{\xi}\right)}^{2\mu_\pm}{\left(\cos{\xi}\right)}^{2\nu_{a,b}}\hypergeom{\alpha_{a,b}^\pm,\beta_{a,b}^\pm;\gamma_\pm;\sin^2{\xi}} \;.\]

Turning back to the $x$ coordinate, the axial solutions in the IR are obtained by identifying \(X_\text{in}^{(1)}(x) \equiv X_{a,b}^+(x)\) (symmetric solutions) and \(X_\text{in}^{(2)}(x) \equiv X_{a,b}^-(x)\) (antisymmetric solutions) in eq.~\eqref{eq:7}, where \(X_{a,b}^\pm(x)\) are obtained after combination of the last equation with eqs.~\eqref{eq:9b} and \eqref{eq:11}:
\begin{align}
    \label{eq:12}
    X_{a,b}^\pm(x)=&{\left(\tanh{x}\right)}^{2\mu_\pm}{\left(\sech{x}\right)}^{2\nu_{a,b}+\frac{1}{2}}\hypergeom{\alpha_{a,b}^\pm,\beta_{a,b}^\pm;\gamma_\pm;\tanh^2{x}} \nonumber \\
    =&{\left(\tanh{x}\right)}^{2\mu_\pm}{\left(\sech{x}\right)}^{2\nu_{a,b}+\frac{1}{2}}\times \nonumber \\
    &\times\hypergeom{\mu_\pm+\nu_{a,b}+\frac{1}{2}\sqrt{\mathcal{E}_{a,b}+\omega_{a,b}},\mu_\pm+\nu_{a,b}-\frac{1}{2}\sqrt{\mathcal{E}_{a,b}+\omega_{a,b}};1\pm\frac{1}{2};\tanh^2{x}} \;.
\end{align}


\subsection{The energy spectrum}
\label{subsec:es}
The energy spectrum is determined by eq.~\eqref{eq:8a}. This is a transcendental equation depending on the ordering-label $a$ and the quantum numbers  \(\mathscr{k}\), \(\ell\) and \(\mathscr{n}\),
where \(\mathscr{n} \in \ZZ_+\). The (\(\mathscr{n}+1\))-th solution corresponds to \(\tilde{E}_{a;\mathscr{k},\ell,\mathscr{n}}\). 
%
%
In order to make it clear, from now on we will include explicitly these indices \footnote{The parameter $b$ is removed for we are considering kinetic-orderings  $b=a$ in this subsection. In sect.\ref{subsec:scpdcm}, $b$ will be considered back}. The eigenstates (orbitals) for each ordering will be denoted by the triple \((\mathscr{k},\ell,\mathscr{n})\).

To carry out the energy calculations and plots, we need to choose the system parameters. For the values $m_1=2$, $m_2=0.5$, \(\delta_x=0.5\) and \(\tilde{V}_0=-50\)  well behaved  axial effective potentials \(\tilde{V}_{\mathscr{p}}\) are guaranteed for the following list of quantum numbers, \(\mathscr{p}_{1,1}\), \(\mathscr{p}_{2,1}\), \(\mathscr{p}_{0,2}\), \(\mathscr{p}_{3,1}\), \(\mathscr{p}_{1,2}\), \(\mathscr{p}_{4,1}\), \(\mathscr{p}_{2,2}\), \(\mathscr{p}_{0,3}\), \(\mathscr{p}_{5,1}\), \(\mathscr{p}_{3,2}\), \(\mathscr{p}_{6,1}\). In Table \ref{tab:1} we can see the energy eigenvalues for the allowed orderings (ZK, MM and BDD). Note that the axial effective potentials with \((\mathscr{k},\ell)=(2,2)\), \((0,3)\), \((5,1)\), \((3,2)\) or \((6,1)\) do not have any bound state, regardless  the ordering. Most of the other pairs of numbers have only one eigenstate (the ground state \(\mathscr{n}=0\)), except for \((\mathscr{k},\ell)=(0,1)\) and \((1,1)\) which have two eigenstates (the ground state and a single excited state). All of the eigenstates are non-degenerated and the trend of energy growth is \((0,1,0) \to (1,1,0) \to (2,1,0) \to (0,2,0) \to (0,1,1) \to (1,1,1) \to (3,1,0) \to (1,2,0) \to (4,1,0)\).

\begin{longtable}{||c||c|c|c||c||}
    \caption{Numerical values of \(\tilde{E}_{a;\mathscr{k},\ell,\mathscr{n}}\) for a PDM particle in a potential given by eq.~\eqref{eq:2}, evaluated for the ZK, MM and BDD orderings.}
    \label{tab:1} \\
    \hline
    \hline
    \boldmath{\((\mathscr{k},\ell,\mathscr{n})\)}   &   \textbf{ZK ordering}    &   \textbf{MM ordering}    &   \textbf{BDD ordering}   &   \textbf{Average}    \\  
    \hline
    \hline
    \boldmath{\((0,1,0)\)}                          &   $-30.566$               &   $-30.183$               &   $-29.805$               &   $-30.185$           \\  
    \hline
    \boldmath{\((0,1,1)\)}                          &   $-3.5175$               &   $-3.5609$               &   $-3.4578$               &   $-3.5120$           \\  
    \hline
    \boldmath{\((1,1,0)\)}                          &   $-21.350$               &   $-20.970$               &   $-20.588$               &   $-20.969$           \\  
    \hline
    \boldmath{\((1,1,1)\)}                          &   $5.4029$                &   $5.4983$                &   $5.7754$                &   $5.5589$            \\  
    \hline
    \boldmath{\((2,1,0)\)}                          &   $-9.2744$               &   $-8.8952$               &   $-8.5048$               &   $-8.8915$           \\  
    \hline
    \boldmath{\((0,2,0)\)}                          &   $-5.0519$               &   $-4.6723$               &   $-4.2782$               &   $-4.6675$           \\  
    \hline
    \boldmath{\((3,1,0)\)}                          &   $5.4775$                &   $5.8596$                &   $6.2656$                &   $5.8676$            \\  
    \hline
    \boldmath{\((1,2,0)\)}                          &   $14.206$                &   $14.594$                &   $15.014$                &   $14.605$            \\  
    \hline
    \boldmath{\((4,1,0)\)}                          &   $22.743$                &   $23.141$                &   $23.581$                &   $23.155$            \\  
    \hline
    \hline
\end{longtable}

\begin{figure}[H]
    \subfigure[\label{fig:3a}]{\includegraphics[width=.32\linewidth]{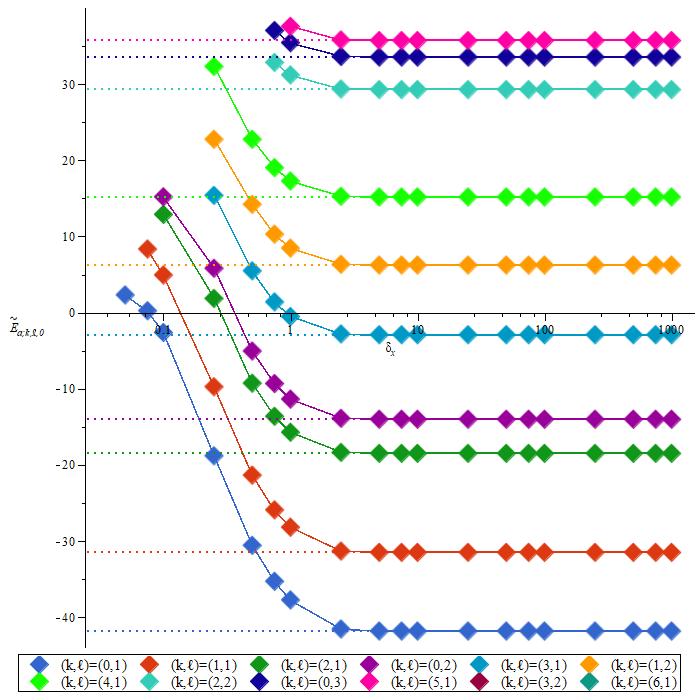}} \ \ 
    \subfigure[\label{fig:3b}]{\includegraphics[width=.32\linewidth]{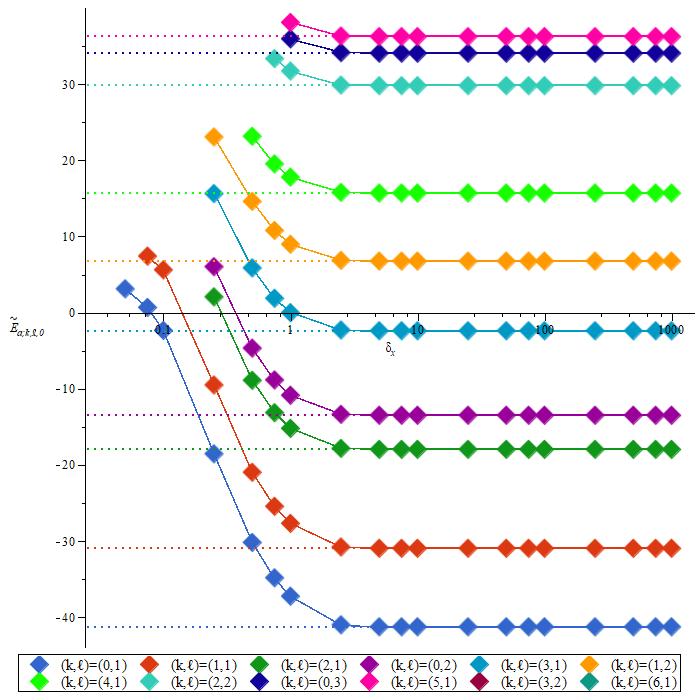}} \ \ 
    \subfigure[\label{fig:3c}]{\includegraphics[width=.32\linewidth]{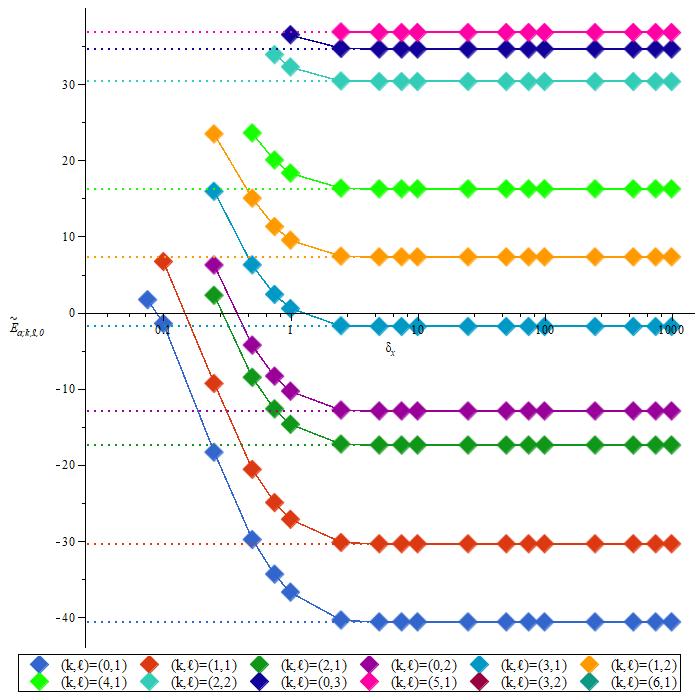}} \ \ 
    \caption{Energy eigenvalues \(\tilde{E}_{a;\mathscr{k},\ell,0}\)  {(vertical axis coordinate)} for the \subref{fig:3a} ZK, \subref{fig:3b} MM and \subref{fig:3c} BDD orderings.  {The horizontal axis coordinate is \(\delta_x\)}.
    System parameters are $m_1=2$, $m_2=0.5$, \(\tilde{V}_0=-50\) and several values of \(\delta_x\) between $0.025$ and $1000$.}
    \label{fig:3}
\end{figure}

We can extract more detailed information about the behavior of the ground state  {for each pair of the quantum numbers \((\mathscr{k},\ell)\) } analyzing how its energy varies with the width of the potential well while keeping the other parameters fixed. The results are shown in Fig.~\ref{fig:3}.


\subsection{Graphics of the axial solutions}
\label{subsec:gas}

\begin{figure}[H]
    \subfigure[\label{fig:4a}]{\includegraphics[width=.32\linewidth]{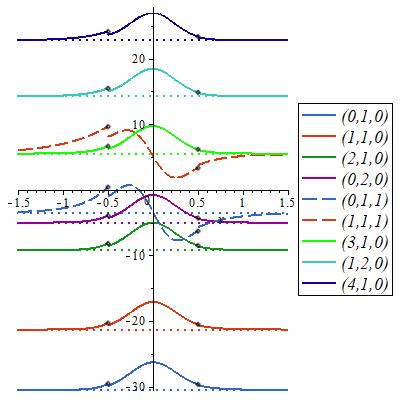}} \ \ 
    \subfigure[\label{fig:4b}]{\includegraphics[width=.32\linewidth]{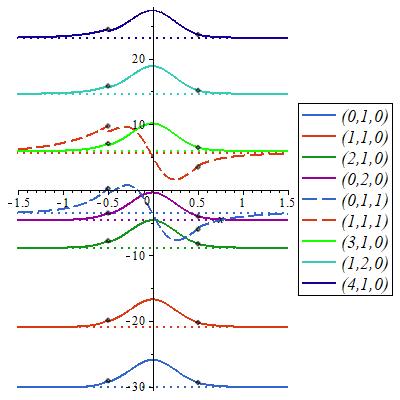}} \ \ 
    \subfigure[\label{fig:4c}]{\includegraphics[width=.32\linewidth]{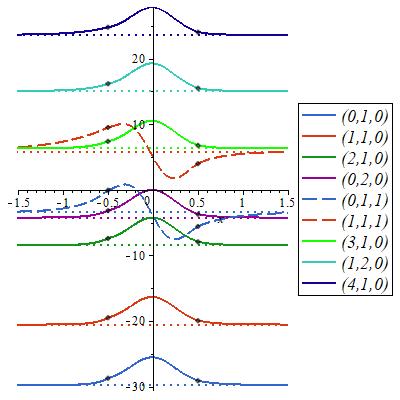}} \ \ 
    \caption{Unnormalized axial solutions of bound states of the \subref{fig:4a} ZK, \subref{fig:4b} MM and \subref{fig:4c} BDD orderings  {(the horizontal axis coordinate is the axial position $x$)}. The amplitude of each solution was appropriately adjusted for a better viewing.}
    \label{fig:4}
\end{figure}

Once the solutions in the IR, \(X_{a,a}^+(x) \equiv X_{a;\mathscr{k},\ell,\mathscr{n}}^+(x)\) and \(X_{a,a}^-(x) \equiv X_{a;\mathscr{k},\ell,\mathscr{n}}^-(x)\), as well as the energy levels, \(\tilde{E}_{a;\mathscr{k},\ell,\mathscr{n}}\), are obtained, eq.~\eqref{eq:7} provides the axial solutions \(X_{a;\mathscr{k},\ell,\mathscr{n}}(x)\), with the coefficients calculated from eq.~\eqref{eq:8b}. The respective graphs are plotted in Fig.~\ref{fig:4} for the orderings allowed.

\begin{figure}[H]
    \subfigure[\label{fig:5a}]{\includegraphics[width=.32\linewidth]{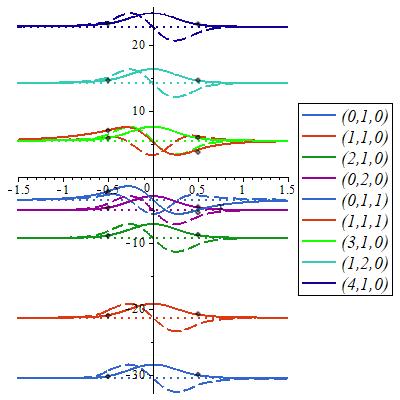}} \ \ 
    \subfigure[\label{fig:5b}]{\includegraphics[width=.32\linewidth]{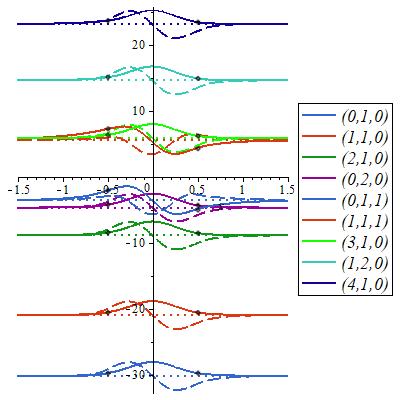}} \ \ 
    \subfigure[\label{fig:5c}]{\includegraphics[width=.32\linewidth]{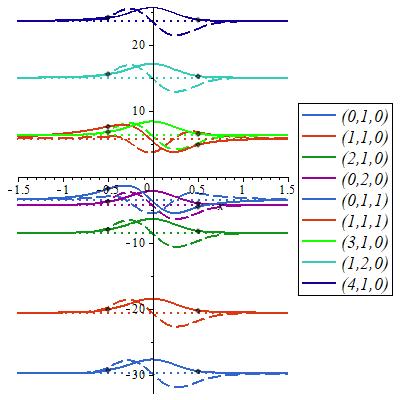}} \ \ 
    \caption{Plots of \(m^aX_{a;\mathscr{k},\ell,\mathscr{n}}(x)\) (solid colored lines) and \(m^{-(a+1)}X'_{a;\mathscr{k},\ell,\mathscr{n}}(x)\) (dashed colored lines) for the \subref{fig:5a} ZK, \subref{fig:5b} MM and \subref{fig:5c} BDD orderings   {(the horizontal axis coordinate is the axial position $x$)}. The amplitude of each solution was appropriately adjusted for a better viewing.}
    \label{fig:5}
\end{figure}

At the boundaries \(x=\pm0.5\) the axial eigenfunctions and corresponding first derivatives can be discontinuous as much as the mass is, but \(m^aX_{a;\mathscr{k},\ell,\mathscr{n}}(x)\) and \(m^{-(a+1)}X'_{a;\mathscr{k},\ell,\mathscr{n}}(x)\) are in fact continuous as we can see in Fig.~\ref{fig:5}.


\subsection{Special case: continuous mass carriers}
\label{subsec:scpdcm}

When we assume PDM without discontinuities, then we can include GW and LK Hamiltonians and consider all the five kinetic-orderings together in the analysis. The eqs.~\eqref{eq:7} and \eqref{eq:12} combined are still valid but we have to take the $b$ parameter into consideration for now \(b\neq a\) is allowed. 

Let us make \(m_1=m_2=\sech^2{\delta_x}\) in eq.~\eqref{eq:2}, so that in Subsection~\ref{subsec:ras:bcas} we must use the continuity of \(X_{a,b}(x)\) and \(X'_{a,b}(x)\) instead of the previous continuity conditions. In this case we have \(\theta_{1;a,b;\mathscr{k},\ell,\mathscr{n}}=\theta_{2;a,b;\mathscr{k},\ell,\mathscr{n}}=\sqrt{\left|\sech^2{\delta_x}\tilde{E}_{a,b;\mathscr{k},\ell,\mathscr{n}}-\mathscr{p}_{\mathscr{k},\ell}^2\right|} \equiv \theta_{a,b;\mathscr{k},\ell\mathscr{n}} .\) Taking into account the parity of the solutions of eq.~\eqref{eq:12}, one has \(\kappa_1^{(j)}={(-1)}^{j+1}\kappa_2^{(j)}\) so that eq.~\eqref{eq:8a} reduces to
\begin{subequations}
    \label{eq:13}
    \begin{equation}
        \label{eq:13a}
        \kappa_{2;a,b;\mathscr{k},\ell,\mathscr{n}}^{(1)}\kappa_{2;a,b;\mathscr{k},\ell,\mathscr{n}}^{(2)}=0 \;,
    \end{equation}
    which gives us the energy spectrum \({\left\{\tilde{E}_{a,b;\mathscr{k},\ell,\mathscr{n}}\right\}}_{\mathscr{n} \in \ZZ_+}\).
    
    Eq.~\eqref{eq:13a} can be divided in two separated equations, one for each parity sector:
    \begin{equation}
        \label{eq:13b}
        \kappa_{2;a,b;\mathscr{k},\ell,2\mathscr{q}+1}^{(1)} \equiv {\left.\diff{X_{a,b;\mathscr{k},\ell,2\mathscr{q}+1}^+}{x}\right|}_{x=\delta_x}+\theta_{a,b;\mathscr{k},\ell,2\mathscr{q}+1}X_{a,b;\mathscr{k},\ell,2\mathscr{q}+1}^+(\delta_x)=0 \;,\quad \mathscr{q} \in \ZZ_+ \;,
    \end{equation}
    \begin{equation}
        \label{eq:13c}
        \kappa_{2;a,b;\mathscr{k},\ell,2\mathscr{q}}^{(2)} \equiv {\left.\diff{X_{a,b;\mathscr{k},\ell,2\mathscr{q}}^-}{x}\right|}_{x=\delta_x}+\theta_{a,b;\mathscr{k},\ell,2\mathscr{q}}X_{a,b;\mathscr{k},\ell,2\mathscr{q}}^-(\delta_x)=0 \;,\quad \mathscr{q} \in \ZZ_+ \;.
    \end{equation}
\end{subequations}
These expressions provide, respectively, the energy eigenvalues associated with antisymmetric, \({\left\{\tilde{E}_{a,b;\mathscr{k},\ell,2\mathscr{q}+1}\right\}}_{\mathscr{q} \in \ZZ_+}\), and symmetric states, \({\left\{\tilde{E}_{a,b;\mathscr{k},\ell,2\mathscr{q}}\right\}}_{\mathscr{q} \in \ZZ_+}\), namely the odd and even eigenstates of the Hamiltonian defined for each combination of $a$ and $b$.

Given that the mass parameter is \(m_1=\sech^2{\delta_x}<1\), the axial effective potentials always have a potential-well shape and \(\mathscr{k}\) and \(\ell\) can take infinitely many values. Let us focus our computation on the first three Bessel zeros, \(\mathscr{p}_{0,1}\), \(\mathscr{p}_{1,1}\) and \(\mathscr{p}_{2,1}\)
for the sake of concision. In Table \ref{tab:3} we show the energy eigenvalues of each of the five orderings.

\begin{longtable}{||c||c|c|c|c|c||c||}
    \caption{Energy eigenvalues \(\tilde{E}_{a,b;\mathscr{k},\ell,\mathscr{n}}\) for a continuous PDM particle in a potential given by eq.~\eqref{eq:2}, evaluated for the GW, ZK, LK, MM and BDD orderings, with \(\delta_x=0.5\) and \(\tilde{V}_0=-50\).}
    \label{tab:3} \\
    \hline
    \hline
    \boldmath{\((\mathscr{k},\ell,\mathscr{n})\)}   &   \makecell{\textbf{GW} \\ \textbf{ordering}}  &    \makecell{\textbf{ZK} \\ \textbf{ordering}} &  \makecell{\textbf{LK} \\ \textbf{ordering}}  &    \makecell{\textbf{MM} \\ \textbf{ordering}} &  \makecell{\textbf{BDD} \\ \textbf{ordering}}   & \textbf{Average}    \\  
    \hline
    \hline
    \boldmath{$(0,1,0)$}    &   $-30.713$ &   $-30.679$ &   $-30.186$ &   $-30.177$ &   $-29.659$ &   $-30.283$ \\  
    \hline
    \boldmath{$(0,1,1)$}    &   $-3.8834$ &   $-3.8028$ &   $-3.3943$ &   $-3.3742$ &   $-2.9080$ &   $-3.4725$ \\  
    \hline
    \boldmath{$(1,1,0)$}    &   $-21.478$ &   $-21.445$ &   $-20.951$ &   $-20.943$ &   $-20.424$ &   $-21.048$ \\  
    \hline
    \boldmath{$(1,1,1)$}    &   $6.1499$  &   $6.2307$  &   $6.6474$  &   $6.6676$  &   $7.1425$  &   $6.5676$  \\  
    \hline
    \boldmath{$(2,1,0)$}    &   $-9.3570$ &   $-9.3241$ &   $-8.8298$ &   $-8.8216$ &   $-8.3027$ &   $-8.9271$ \\  
    \hline
    \boldmath{$(2,1,1)$}    &   $19.280$  &   $19.361$  &   $19.787$  &   $19.808$  &   $20.292$  &   $19.706$  \\  
    \hline
    \hline
\end{longtable}

Note that, for each pair \((\mathscr{k},\ell)\) among the three here analyzed, there are two bound states. Again, the eigenstates are non-degenerated and the energy growth as \((0,1,0) \to (1,1,0) \to (2,1,0) \to (0,1,1) \to (1,1,1) \to (2,1,1)\).

\begin{figure}[H]
    \subfigure[\label{fig:6a}]{\includegraphics[width=.19\linewidth]{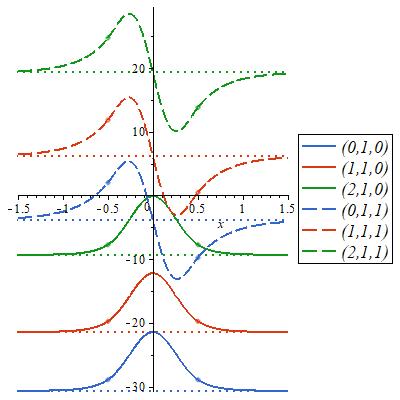}} \
    \subfigure[\label{fig:6b}]{\includegraphics[width=.19\linewidth]{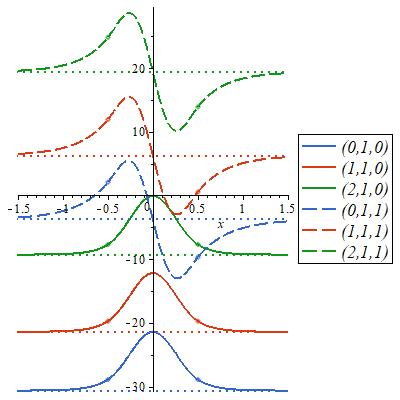}} \
    \subfigure[\label{fig:6c}]{\includegraphics[width=.19\linewidth]{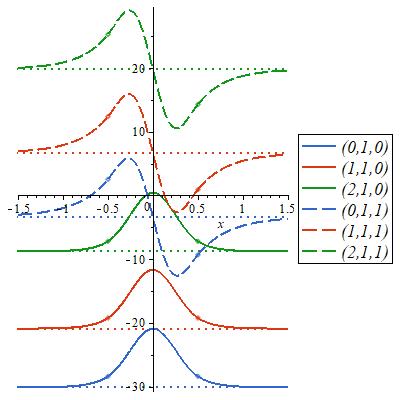}} \
    \subfigure[\label{fig:6d}]{\includegraphics[width=.19\linewidth]{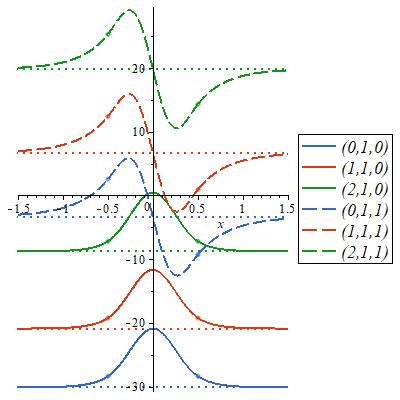}} \
    \subfigure[\label{fig:6e}]{\includegraphics[width=.19\linewidth]{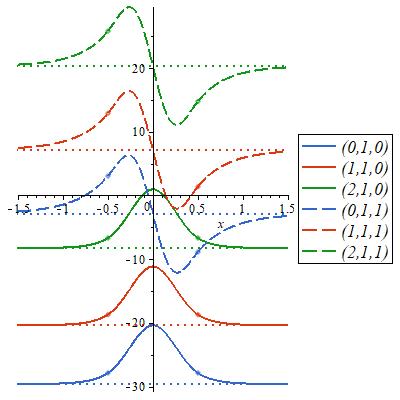}}
    \caption{Unormalized bound-states of the \subref{fig:6a} GW, \subref{fig:6b} ZK, \subref{fig:6c} LK, \subref{fig:6d} MM and \subref{fig:6e} BDD orderings, for \((\mathscr{k},\ell)=(0,1)\), $(1,1)$ and $(2,1)$  {(the horizontal axis coordinate is the axial position $x$)}. The amplitude of each solution was appropriately adjusted for a better viewing.}
    \label{fig:6}
\end{figure}

The plots for the axial eigenfunctions with the energies of Table~\ref{tab:3} are shown in Fig.~\ref{fig:6}.


\section{Eigenstates}
\label{sec:e}

The set of eqs.~\eqref{eq:5}, \eqref{eq:6}, \eqref{eq:7}, \eqref{eq:8} and \eqref{eq:12}  give us the complete spectrum of bounded eigenstates \(\psi_{a;\mathscr{k},\ell,\mathscr{n}}(\boldsymbol{r})\):
\begin{subequations}
    \label{eq:14}
    \begin{equation}
        \label{eq:14a}
        \psi_{a;\mathscr{k},\ell,\mathscr{n}}(\boldsymbol{r})=\mathfrak{C}_{a;\mathscr{k},\ell,\mathscr{n}}e^{i\mathscr{k}\phi}J_\mathscr{k}\!\left(\textcolor{red}{\mathscr{p}_{\mathscr{k},\ell}}r\right)\mathcal{X}_{a;\mathscr{k},\ell,\mathscr{n}}(x)
    \end{equation}
     {inside the cylinder and zero outside,} where \(\mathfrak{C}_{a;\mathscr{k},\ell,\mathscr{n}} \equiv A_\mathscr{k}B_{\mathscr{k},\ell}C_{a;\mathscr{k},\ell,\mathscr{n}}\) results from normalization of \(\psi_{a;\mathscr{k},\ell,\mathscr{n}}(\boldsymbol{r})\):
    \begin{equation}
        \label{eq:14b}
        \mathfrak{C}_{a;\mathscr{k},\ell,\mathscr{n}}=\frac{1}{\sqrt{2\pi}}{\left[\int_0^{\textcolor{red}{\delta_r}}{{J_\mathscr{k}\!\left(\textcolor{red}{\mathscr{p}_{\mathscr{k},\ell}}r\right)}^2r\text{d}r}\int_{-\infty}^\infty{{\mathcal{X}_{a;\mathscr{k},\ell,\mathscr{n}}(x)}^2\text{d}x}\right]}^{-1/2} \;.
    \end{equation}
\end{subequations}

\begin{figure}[H]
    \includegraphics[width=\textwidth]{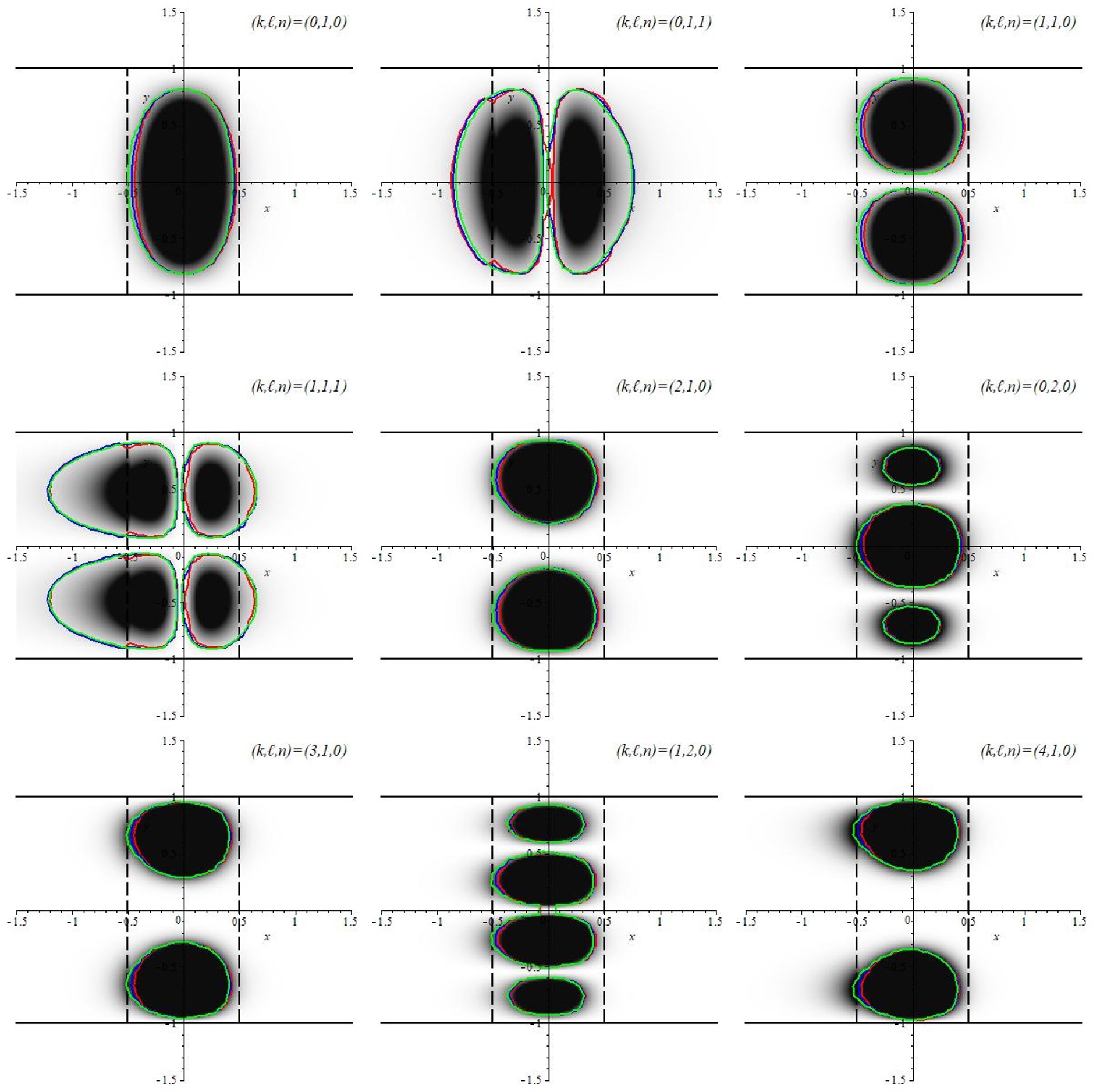}
    \caption{Density plots of \(\Bar{\rho}_{\mathscr{k},\ell,\mathscr{n}}\) (dark zones are associated with a higher probability density). The contour lines of constant \(\rho_{a;\mathscr{k},\ell,\mathscr{n}}\) referring to the ZK, MM and BDD orderings are in red, blue and green, respectively. The two solid black horizontal lines indicate the cylindrical region to which the particle is radially confined and the two dashed black vertical lines mark the IR.}
    \label{fig:7}
\end{figure}

\begin{figure}[H]
    \includegraphics[width=\textwidth]{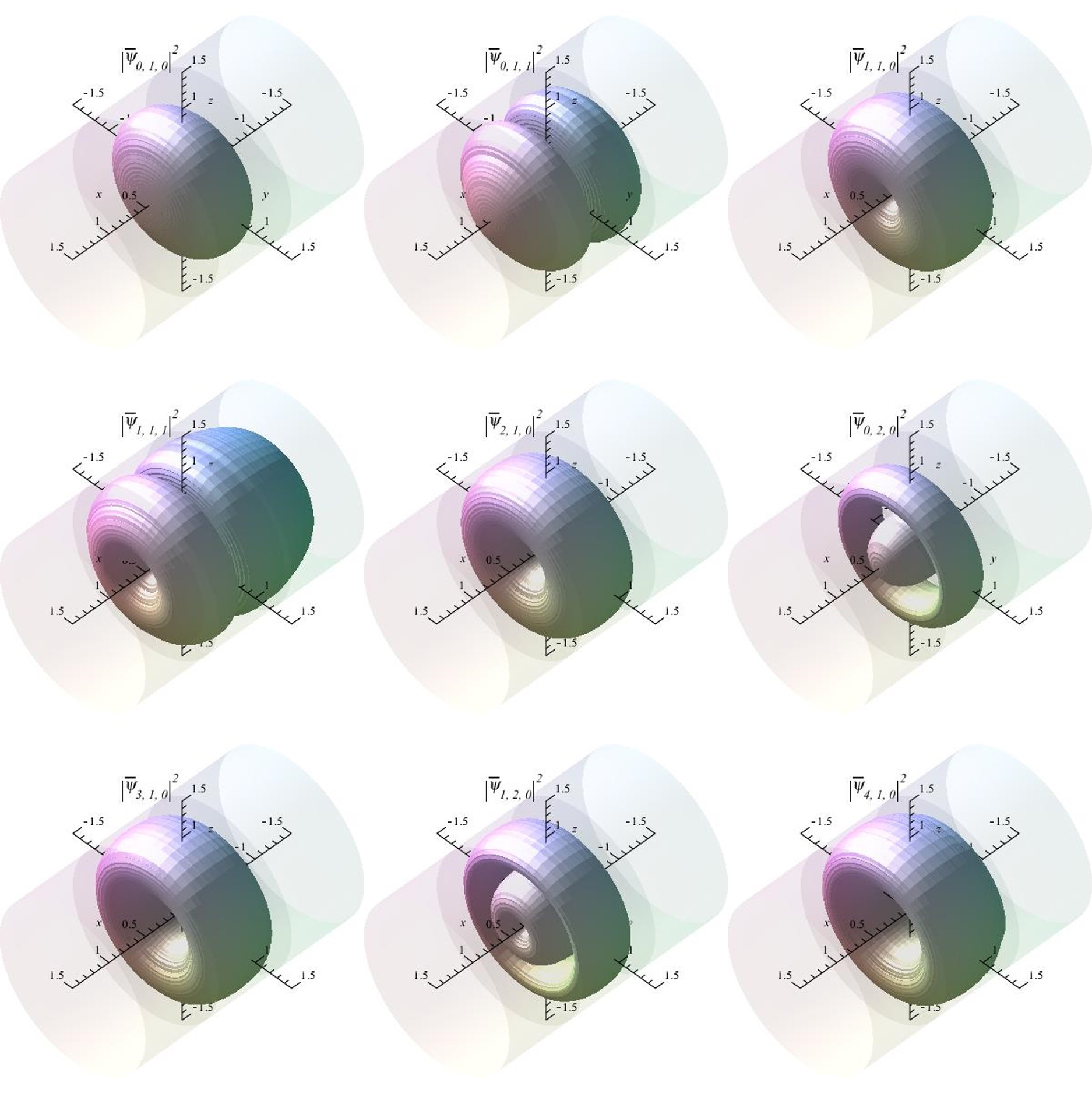}
    \caption{Surfaces of constant \(\Bar{\rho}_{\mathscr{k},\ell,\mathscr{n}}\) (we can identify the cylindrical region and the interfaces of the IR by the transparent surfaces surrounding the graphics).}
    \label{fig:8}
\end{figure}

From eqs.~\eqref{eq:14} we can write the probability density for each eigenstate as
\[\rho_{a;\mathscr{k},\ell,\mathscr{n}}(\boldsymbol{r}) \equiv \sqabs{\psi_{a;\mathscr{k},\ell,\mathscr{n}}(\boldsymbol{r})}=\frac{1}{2\pi}\frac{{J_\mathscr{k}\!\left(\textcolor{red}{\mathscr{p}_{\mathscr{k},\ell}}r\right)}^2{\mathcal{X}_{a;\mathscr{k},\ell,\mathscr{n}}(x)}^2}{\int_0^\infty{{J_\mathscr{k}\!\left(\textcolor{red}{\mathscr{p}_{\mathscr{k},\ell}}\mathscr{r}\right)}^2\mathscr{r}\text{d}\mathscr{r}}\int_{-\infty}^\infty{{\mathcal{X}_{a;\mathscr{k},\ell,\mathscr{n}}(\mathscr{x})}^2\text{d}\mathscr{x}}} \;.\]
In Figs.~\ref{fig:7} and \ref{fig:8} we plot the graphs of the \emph{average} probability density of the three orderings ZK, MM and BDD, represented by \(\Bar{\rho}_{\mathscr{k},\ell,\mathscr{n}} \equiv \sqabs{\Bar{\psi}_{\mathscr{k},\ell,\mathscr{n}}}\) (note that since only the square of eigenfunctions has physical meaning, we do not take into account the negative values of \(\mathscr{k}\) \cite[eq.~9.1.5]{abramowitz:stegun:1970}).

\section{Conclusions and final remarks}
\label{sec:c}
In this work we have obtained the exact bound solutions of the generalized three-dimensional position-dependent mass  Schrodinger equation for a cylindrical quantum dot. The energy spectrum depends on three quantum numbers (\(\mathscr{k}\), \(\ell\) and \(\mathscr{n}\)) and two ordering-labels ($a,b$) which characterize the Hamiltonian adopted. We have presented a comprehensive model of a generic nano-heterostructure. Within such quantum cylinder, the particle mass and the potential are of hyperbolic type in the region between two uniform different layers. Boundary conditions have been taken both open and closed in radial and axial directions.


Our spectral values have been calculated for a significant set of states in an expressive number of cases. We have found that the energy spectrum grows smoothly, but strictly, as a function of the kinetic-Hamiltonian adopted to design the structure: For discontinuous PDM models it grows according to the choices \(\text{ZK} \to \text{MM} \to \text{BDD}\). For continuous PDM models the growing sequence is \(\text{GW} \to \text{ZK} \to \text{LK} \to \text{MM} \to \text{BDD}\). This awareness may allow the appropriate choice of model to better fit experimental data.  

Another feature that we have noticed is the absence of bound-states below specific values of \(\delta_x\), see Fig.~\ref{fig:3}. For example, for \((\mathscr{k},\ell)=(3,2)\) or $(6,1)$ there are simply no bound states for any \(\delta_x\). This can be explained by the fact that the effective potential-well is very shallow in theses cases (the depths are respectively $2.3612$ and $0.63686$). Fig.~\ref{fig:3} also reveals that as \(\delta_x\) grows, the energies of the ground states decrease asymptotically up to a threshold that depends on $a$, \(\mathscr{k}\) and \(\ell\). This means that variations in the width of wide IR have little effect on the energy levels. Indeed, we have verified that letting \(\delta_x \to \infty\), the axial effective-potential tends to an infinite potential-well and the energy spectrum is given by from \(\Xi_{a,a;\mathscr{k},\ell}^\pm(\pm\pi/2)=0\). 
For \(\omega_{a,a;\mathscr{k},\ell} \geq 0\), corresponding to the discontinuous case,  this equation results in an analytical expression  for \(\tilde{E}_{a;\mathscr{k},\ell,\mathscr{n}}\)  {(see \cite{christiansen:cunha:2013, cunha:christiansen:2013, christiansen:cunha:2014, lima:christiansen:2022} and \cite[eq.~15.1.20]{abramowitz:stegun:1970})}
\[{\left.\tilde{E}_{a;\mathscr{k},\ell,\mathscr{n}}\right|}_{ {\delta_x\to\infty}}={\left(\mathscr{n}+1-2\nu_{a,a;\mathscr{k},\ell}\right)}^2-\omega_{a,a;\mathscr{k},\ell}+V_{a,a;\mathscr{k},\ell}^{(0)} \;.\]
Using Table~\ref{tab:2} we checked the consistence of the results and verified that \(\delta_x=1000\) is a good approximation (up to four significant digits) for an infinite potential-well (in such a case  \((\mathscr{k},\ell)=(3,2)\) and $(6,1)$ would then allow bound states).
\begin{longtable}{||c||c|c|c||c|c|c||}
    \caption{Comparison between the  \(\mathscr{n}=0\) states obtained from eq.~\eqref{eq:8a} for \(\delta_x=1000\) and those one from the analytic expression for \(\delta_x \to \infty\).}
    \label{tab:2} \\
    \cline{2-7}
    \noalign{\vskip\doublerulesep\vskip-\arrayrulewidth}
    \cline{2-7}
    \multicolumn{1}{c||}{} & \multicolumn{3}{c||}{ {\(\delta_x=1000\)}} & \multicolumn{3}{c||}{ {\(\delta_x \to \infty\)}}  \\ 
    \hline
    \hline
    \boldmath{\((\mathscr{k},\ell)\)} & \makecell{\textbf{ZK} \\ \textbf{ordering}} & \makecell{\textbf{MM} \\ \textbf{ordering}} & \makecell{\textbf{BDD} \\ \textbf{ordering}} & \makecell{\textbf{ZK} \\ \textbf{ordering}} & \makecell{\textbf{MM} \\ \textbf{ordering}} & \makecell{\textbf{BDD} \\ \textbf{ordering}}  \\ 
    \hline
    \hline
    \boldmath{$(0,1)$}  &   $-41.812$	&   $-41.261$   &	$-40.612$   &   $-41.812$   &	$-41.261$   &	$-40.613$   \\  
    \hline
    \boldmath{$(1,1)$}  &   $-31.486$	&   $-30.954$   &	$-30.358$   &   $-31.487$   &	$-30.954$   &	$-30.358$   \\  
    \hline
    \boldmath{$(2,1)$}  &   $-18.490$	&   $-17.965$   &	$-17.393$   &   $-18.490$   &	$-17.964$   &	$-17.394$   \\  
    \hline
    \boldmath{$(0,2)$}  &   $-14.009$	&   $-13.486$   &	$-12.919$   &   $-14.010$   &	$-13.482$   &	$-12.918$   \\  
    \hline
    \boldmath{$(3,1)$}  &   $-2.9134$	&   $-2.3938$   &	$-1.8355$   &   $-2.9160$   &	$-2.3900$   &	$-1.8360$   \\  
    \hline
    \boldmath{$(1,2)$}  &   $6.2340$    &	$6.7518$    &	$7.3050$    &   $6.2340$    &	$6.7550$    &	$7.3030$    \\  
    \hline
    \boldmath{$(4,1)$}  &   $15.171$    &	$15.688$    &	$16.237$    &   $15.172$    &	$15.691$    &	$16.238$    \\  
    \hline
    \boldmath{$(2,2)$}  &   $29.267$    &	$29.782$    &	$30.326$    &   $29.263$    &	$29.781$    &	$30.326$    \\  
    \hline
    \boldmath{$(0,3)$}  &   $33.541$    &	$34.055$    &	$34.598$    &   $33.546$    &	$34.052$    &	$34.604$    \\  
    \hline
    \boldmath{$(5,1)$}  &   $35.710$    &	$36.225$    &	$36.767$    &   $35.713$    &	$36.230$    &	$36.769$    \\  
    \hline
    \boldmath{$(3,2)$}  &   –           &	–           &	–	        &   $55.040$    &	$55.560$    &	$56.090$    \\  
    \hline
    \boldmath{$(6,1)$}  &   –           &	–           &	–	        &   $58.660$    &	$59.160$    &	$59.710$    \\  
    \hline
    \hline
\end{longtable}

\begin{figure}[H]
    \includegraphics[width=.80\linewidth]{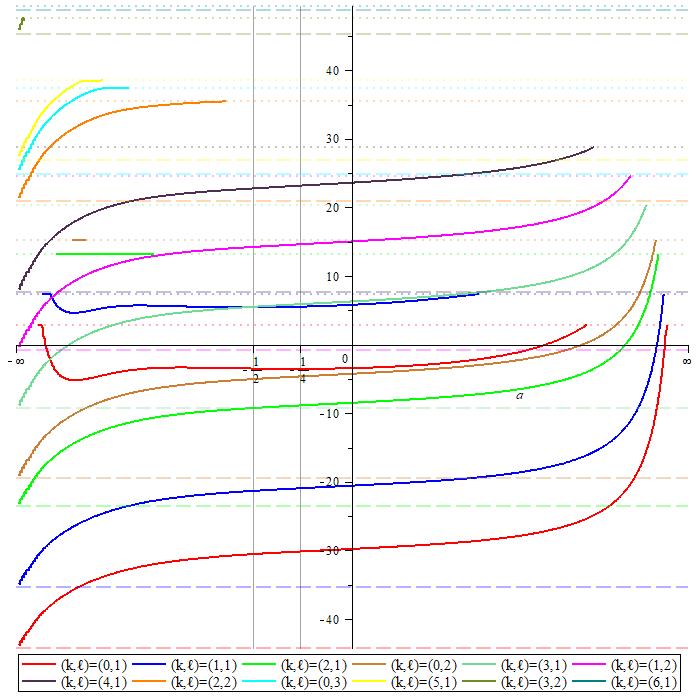}
    \caption{
     {
    Plot of bound eigenenergies \(\tilde{E}_{a;\mathscr{k},\ell,\mathscr{n}}\) as a function of $a$ (parameters are fixed as before). Vertical gridlines are for the ZK, MM and BDD orderings, while the horizontal colored lines are for the minimum (dashed lines) and maximum (dotted lines) of the axial effective potential well. The horizontal axis has been rescaled in order to show the entire domain. Each color represents a different  \((\mathscr{k},\ell)\) pair as indicated; two curves with same color differ by the quantum number \(\mathscr{n}\): the one positioned higher indicates \(\mathscr{n}=1\) and the one positioned lower indicates \(\mathscr{n}=0\) ---  e.g. the two red lines, are for the orbitals      \((\mathscr{k},\ell,\mathscr{n})=(0,1,0)\) and $(0,1,1)$; the two blue lines are for \((\mathscr{k},\ell,\mathscr{n})=(1,1,0)\) and $(1,1,1)$.
    }
    }
    \label{fig:9}
\end{figure}

 {The detailed analysis of the energy eigenvalues as a function of the ordering parameter $a$ (see Fig.~\ref{fig:9}) shows that they  do not differ significantly for ZK, MM and BDD choices. A general comparison indicates that 
}
 {
i) the eigenenergies of the \(\mathscr{n}=0\) states of the system grow monotonically, as a function of $a$, from the minimum of the \(\mathscr{p}\)-depending axial effective potential \(\tilde{V}_{{\mathscr{p}_{\mathscr{k},\ell}}_{\text{min}}}(x) = \tilde{V}_0+\mathscr{p}_{\mathscr{k},\ell}^2\) ; ii) the curves are almost flat in the region between the ZK, MM and BDD orderings (\(-\frac{1}{2} \leq a \leq 0\)) meaning little difference among these orderings; iii) there is a maximum value of $a>0$ for which bound states exist; iv) new states emerge in some intervals outside the ZK, MM and BDD orderings---e.g. \((\mathscr{k},\ell,\mathscr{n})=(2,1,1)\) (upper green line), $(0,2,1)$ (upper golden line), $(2,2,0)$ (coral line), $(0,3,0)$ (cyan line), $(5,1,0)$ (yellow line) and $(3,2,0)$ (olive color line) (for \((\mathscr{k},\ell)=(6,1)\) there is no bound state for any value of $a$).
}

 {
Our model results can be useful in the general understanding and design of semiconductor devices and quantum dots provided that the parameters are properly adjusted. In semiconducting nanowires and similar applications , see e.g. \cite{voigtlander:2005, zhuang:etal:2009, rigutti:2014, zhu:2017, kasapoglu:etal:2022}, one should fix the scale parameter \(\epsilon \sim \qty{e{-9}}{\metre}\) 
and  the energy levels \(E_{a;\mathscr{k},\ell,\mathscr{n}}\) are given by the above computed \(\tilde{E}_{a;\mathscr{k},\ell,\mathscr{n}}\) multiplied by \(\frac{\hbar^2}{2\epsilon^2m_0} \approx \frac{0.038100}{m_0/m_e}\,\unit{\eV}\) where \(m_e\) is the free electron mass 
 (for microwires, see e.g. \cite{ioysher:kantser:leporda:1997},  \(\epsilon \sim \qty{e{-6}}{\metre}\) and this factor results  \(\sim 10^6\) times smaller). Based on works such as \cite{dang:neu:romestain:1982, fanciulli:lei:moustakas:1993, miwa:fukumoto:1993, fan:etal:1996, li:etal:1996, long:harrison:hagston:1996, paiva:etal:2002, margaritondo:2005, harrison:valavanis:2016} it is possible to estimate different values of $m_0/m_e$ going from order $10^{-2}$ to $10$ and even more  \cite{wasserman:2005}. If we use our model to describe  a \emph{pn} junction (see, e.g.  \cite{sah:noyce:shockley:1957} and \cite[p.~366]{poole:darwazeh:2015}), we set \(\delta_x \sim 10^2\) to $10^3$ in order to fit the size of the intermediate region, IR, to the depletion zone, typically of order \(\qty{e{-6}}{\metre}\) \cite[p.~367]{poole:darwazeh:2015}.
}

 {
Being in possession of the exact eigenfunctions, \(\Psi_{a;\mathscr{k},\ell,\mathscr{n}}(\textbf{r}) = \epsilon^{-3/2}\psi_{a;\mathscr{k},\ell,\mathscr{n}}(\textbf{r}/\epsilon)\), and eigenvalues, \(E_{a;\mathscr{k},\ell,\mathscr{n}} = \left( \frac{0.038100}{m_0/m_e}\,\unit{\eV} \right) \tilde{E}_{a;\mathscr{k},\ell,\mathscr{n}}\), of the carriers in the cylindrical nanotube we are able to compute its quantum properties. For example, there are several papers (see e.g. \cite{karabulut:baskoutas:2008, xie:2010, miranda:mora-ramos:duque:2013, kasapoglu:etal:2021, kasapoglu:etal:2022, yucel:kasapoglu:duque:2022, sakiroglu:yucel:kasapoglu:2023}) dedicated to analyzing the optical behavior of quantum dots by calculating the absorption coefficient, \(\zeta\), and the relative changes at the same order of the refractive index, \( {\Delta n}/{n}\), for an incident linearly polarized light beam. For a transition \((\mathscr{k}_i,\ell_i,\mathscr{n}_i) \to (\mathscr{k}_j,\ell_j,\mathscr{n}_j)\), we can compute such quantities (at first order for simplicity) as functions of the energy \(E_\gamma\) of the incident photon \cite{kasapoglu:etal:2021} by using a density matrix approach and a perturbation expansion method \cite{unlu:karabulut:safak:2006},
\[\zeta_{a;ij}\left(E_\gamma\right) = \sqrt{\frac{\mu_0}{\varepsilon}} 
\frac{\sigma_v\Gamma_{ij}\sqabs{\mathcal{M}_{a;ij}}\;\;{E_\gamma}/{\hbar}}
{{\left({\Delta E}_{a;ij}-E_\gamma\right)}^2+\Gamma_{ij}^2 } ,
\;\;\;\;\; \text{and} \;\;\;\;\;
\frac{{\Delta n}_{a;ij}\left(E_\gamma\right)}{n_r} = \frac{\sigma_v\sqabs{\mathcal{M}_{a;ij}}}{2\varepsilon} \frac{{\Delta E}_{a;ij}-E_\gamma}{{\left({\Delta E}_{a;ij}-E_\gamma\right)}^2+\Gamma_{ij}^2} \]
Here, \(\mu_0
\) is the permeability of the vacuum, \(\varepsilon = \varepsilon_0n_r^2 = \varepsilon_0\varepsilon_r\) is the real part of the permittivity of the material (\(\varepsilon_0
\) is the permittivity of the vacuum, \(\varepsilon_r\) is the static dielectric constant of the material and \(n_r = \sqrt{\varepsilon_r}\) is the refractive index of the medium), \(\sigma_v\) is the carrier density, \(\Gamma_{ij} = \hbar/T_{ij}\) is the relaxation rate---is a damping-related Lorentzian term associated with exciton scattering losses in the system \cite{miranda:mora-ramos:duque:2013}---($T_{ij}$ is the time relaxation), \(\mathcal{M}_{a;ij} \equiv \langle \Psi_{a;\mathscr{k}_j,\ell_j,\mathscr{n}_j} \left| ex \right| \Psi_{a;\mathscr{k}_i,\ell_i,\mathscr{n}_i} \rangle\) is the dipole matrix element for $x$-polarized incident radiation (\(e
\) is the elementary charge) and \({\Delta E}_{a;ij} \equiv E_{a;\mathscr{k}_j,\ell_j,\mathscr{n}_j}-E_{a;\mathscr{k}_i,\ell_i,\mathscr{n}_i}\) is the energy gap between the levels.
}

 {
For \(\text{Al}_\chi\text{Ga}_{1-\chi}\text{As}\) typical values are \(m_0m_1 = M_{\text{GaAs}} = 0.067m_e\), \(\varepsilon_r = n_r^2 = 13.0\) (the {GaAs} static dielectric constant), \(\sigma_v = \qty{3.0e22}{\per \metre\tothe{3}}\) and \(\Gamma_{ij} = \qty{6.6}{meV}\). After the evaluation of the dipole matrix element with the eigenfunctions in eq.~\eqref{eq:14} we use these parameters in the above equations to compare \(\zeta\) and \(\Delta n/n\) as functions of \(E_\gamma\) for each ordering and show that they peak at different values of the energy transition. Since the first four states are symmetric in the $x$ variable (see Fig.~\ref{fig:4}), the dipole matrix element for the transitions between these states is zero, so in Fig.~\ref{fig:10} are shown the results for the transition \((0,1,0) \to (0,1,1)\) (i.e., between the ground state and the fourth excited state).
\begin{figure}[H]
	\subfigure[\label{fig:10a}]{\includegraphics[width=.49\linewidth]{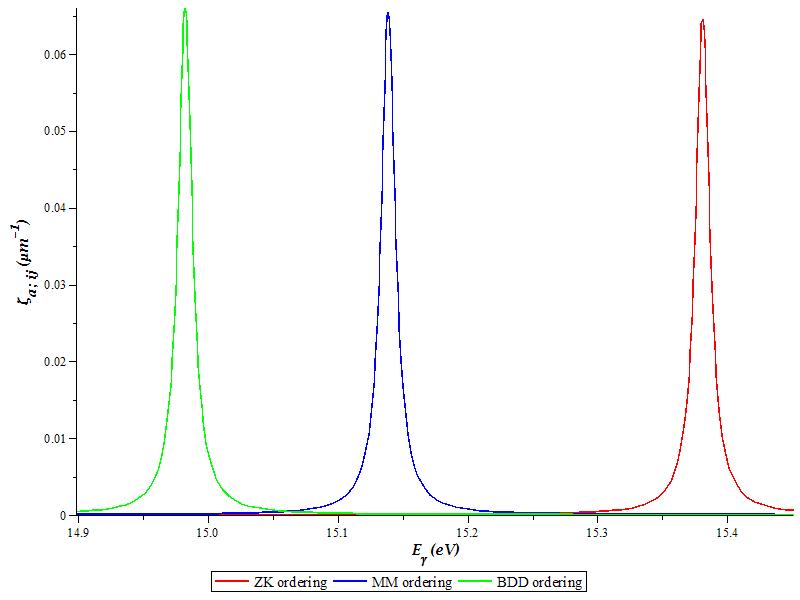}} \ \
	\subfigure[\label{fig:10b}]{\includegraphics[width=.49\linewidth]{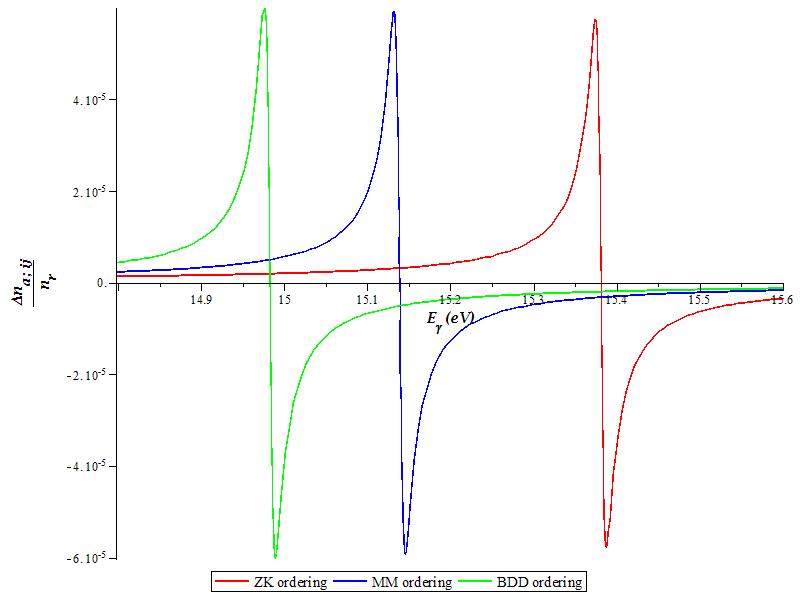}} 
    \caption{
     {
    Comparison between the three orderings of \subref{fig:10a} absorption coefficient and \subref{fig:10b} relative refractive index.
    }
    }
	\label{fig:10}
\end{figure}
}

\section*{Acknowledgement}
The authors thank Conselho Nacional de Desenvolvimento Científico e Tecnológico (CNPq) for partial support.


\bibliographystyle{ieeetr}
\bibliography{refs}

\end{document}